\documentclass[preprintnumbers,nofootinbib]{revtex4}
\usepackage{graphicx}% Include figure files
\usepackage{dcolumn}% Align table columns on decimal point
\usepackage{bm}% bold math
\usepackage{epsfig,amsmath}
\usepackage{amssymb}
\usepackage{subfigure}
\usepackage{float}
\usepackage{ulem}
\usepackage{comment}
\usepackage[usenames,dvipsnames]{color}
\usepackage[pagebackref=false, colorlinks=true]{hyperref}
\usepackage[labelfont=bf]{caption}
\definecolor{redish}{rgb}{0.7,0.2,0.0}  % color defined in (r=red,g=green,b=blue) model
\definecolor{bluish}{rgb}{0.2,0.5,0.8}

\makeatletter
\renewcommand{\fnum@figure}{Figure. \thefigure}
\makeatother

\hypersetup{linkcolor=redish,          % color of internal links
                  citecolor=blue,        % color of links to bibliography
                  filecolor=magenta,      % color of file links
                  urlcolor=bluish}          % color of the url

\DeclareFontFamily{U}{rsfs}{}         % Formal Script            %
\DeclareFontShape{U}{rsfs}{m}{n}{<5> rsfs5 <6><7> rsfs7          %
  <8><9><10><10.95><12><14.4><17.28><20.74><24.88> rsfs10}{}     %
\DeclareMathAlphabet{\mathfs}{U}{rsfs}{m}{n}

\begin{document}

\title{Spin Precession in magnetized Kerr spacetime}

\author{Karthik Iyer}
\email{karthik.mcnsmpl2023@learner.manipal.edu} 
\affiliation{Manipal Centre for Natural Sciences, Manipal Academy of Higher Education, Manipal 576104, Karnataka, India}

\author{Chandrachur Chakraborty}
\email{chandrachur.c@manipal.edu} 
\affiliation{Manipal Centre for Natural Sciences, Manipal Academy of Higher Education, Manipal 576104, Karnataka, India}

\begin{abstract}
We present an exact analytical investigation of spin precession for a test gyroscope in the magnetized Kerr spacetime—an exact electrovacuum solution to the Einstein-Maxwell equations. Our approach accommodates arbitrary magnetic field strengths, enabling a unified treatment across both weak and ultra-strong field regimes. The analysis reveals distinct spin precession behaviors near rotating collapsed objects, which differ characteristically between black holes and naked singularities, offering a potential observational means to differentiate them. The external magnetic field induces a nontrivial modification of the precession frequency through its interaction with the spacetime's gravitoelectromagnetic structure. In the weak-field limit, magnetic fields generally reduce the precession rate, though the effect depends sensitively on the motion and orientation of the test gyro close to the collapsed object. As a special case, we show that in the presence of magnetic fields, the spin precession frequency due to gravitomagnetic effect acquires a long-range $1/r$ (where $r$ is the distance from the central object to the test gyro) correction in contrast to the standard $1/r^3$ falloff. In addition, we obtain the exact geodetic precession (gravitoelectric effect) frequency for a gyroscope in magnetized Schwarzschild spacetime, showing that the magnetic field enhances ($\propto r^{1/2}$) geodetic precession in contrast to the standard $1/r^{5/2}$ falloff. Our results provide observationally testable predictions relevant for black holes in strong magnetic environments, including those possibly realized near magnetars or in the early universe. In particular, the strong-field behavior of spin precession could have important implications for transmuted black holes formed via collapse or mergers of magnetized progenitors in both astrophysical and cosmological contexts.
\end{abstract}

\maketitle

\section{\label{intro}Introduction}
The presence of magnetic fields around black holes (BHs) is no longer just a theoretical possibility but an observational reality. Recent results from the Event Horizon Telescope (EHT) have revealed the polarized emission surrounding the supermassive black hole (SMBH) M87*, providing an estimate of the average magnetic field strength in the range of $\sim 1–30$ G on event horizon scales \cite{EHT7,EHT8}. Observations of Sagittarius A* further suggest that magnetic fields in the range of $30-100$ G are necessary to account for the observed synchrotron radiation near the event horizon \cite{eatough2013strong}. Additionally, studies of the X-ray corona in Cygnus X-1 indicate the presence of even stronger magnetic fields, reaching  $10^5-10^7$ G \cite{santomnrs}. These findings provide compelling evidence that BHs are typically immersed in nonzero magnetic fields.

\vspace{2mm}

Magnetic fields play a crucial role in explaining various astrophysical phenomena, including magnetohydrodynamics (MHD) simulations of accretion mechanisms \cite{Gammie_2004}, constraints on theoretical interpretations of ring structures and relativistic jet-launching mechanisms \cite{EHT7, EHT8}, gravitational Larmor precession \cite{CCGLP, GaltsovBHMag}, the gravitational Meissner effect \cite{GME}, and the Blandford–Znajek process \cite{BZog}, among others. Recent studies have increasingly focused on more astrophysically realistic scenarios, particularly Kerr BHs immersed in uniform magnetic fields. These investigations have uncovered significant effects such as logarithmic correction to the distance between the source and the observer in gravitational Faraday rotation and the gravitational Stern-Gerlach effect for the said spacetime as direct consequences of the presence of magnetic fields \cite{CCFaraday}. A recent study \cite{CCPenrose} demonstrated that the efficiency of the magnetic Penrose process (MPP) depends strongly on the magnetic field strength, spanning from weak to ultra-strong regimes that can substantially deform the Kerr spacetime. {Magnetic fields also play a crucial role in shaping accretion disk dynamics and give rise to several observable astrophysical signatures \cite{Prasanna1978,Wiita1983,Iyer1985}. The motion of test particles in axisymmetric spacetimes with scalar and electromagnetic fields has been recently revisited in \cite{Horak2025}, highlighting the influence of these fields on orbital structure and stability. Foundational studies, beginning with Wald’s work \cite{WaldBHmag}, and followed by detailed analyses \cite{bini, mbo, IyerVishveshwara, saadati}, have demonstrated how external magnetic fields affect charged particle motion, orbital stability, and accretion in Kerr or magnetized black hole spacetimes. More recent investigations \cite{mo} have extended these ideas, focusing on energy extraction mechanisms and magnetic interactions in strong gravity environments. While such studies have significantly advanced our understanding of charged particle dynamics, the evolution of test spin in magnetized, rotating backgrounds remains comparatively underexplored.} Inspired by these recent developments, the study of spin precession in magnetized Kerr spacetime assumes significant importance, particularly in understanding the motion of a test spinning particle in rotating spacetimes with substantial magnetic fields.

\vspace{2mm}

A new era is now opening up for testing GR's validity in the strong gravitational regime, which is the most exciting and fascinating regime. One of the key predictions of GR in this regime is that the spin of a freely falling gyroscope undergoes precession due to two distinct effects. The first is geodetic precession, also known as de Sitter (dS) precession, which arises purely from spacetime curvature \cite{SakinaSch,SakinaKerr,KroriSpin}. On the other hand, if the spinning test object (gyroscope) moves in a stationary, axisymmetric spacetime such as that of a rotating black hole, it experiences an additional precession known as Lense-Thirring (LT) precession, caused by the dragging of inertial frames. Unlike geodetic precession, LT precession persists even when the gyroscope follows a non-geodesic trajectory \cite{StraumannGR,Chakraborty2014,CCPD,CCNeutronStar,Chakraborty2015Anomalous,Chatterjee_2017}. Therefore, obtaining the total precession frequency of a test gyroscope requires accounting for all these effects. Geodetic and LT precession has been efficiently treated within the framework of gravitoelectromagnetism (GEM) formalism \cite{mashhoonGEM,mashhoontime,mashhoonLT,biniGIGEM} and the references therein \footnote{Geodetic precession is called as \textit{gravitoelectric} (GE) effect, while LT precession is called as \textit{gravitomagnetic} (GM) effect.}. In fact, the precise measurement of this field via gyroscopes in a drag-free satellite about the Earth was measured by Gravity probe B whose findings revealed that the geodetic precession rate was $6,601.8 \pm 18.3$ milliarcseconds per year (mas/yr) and the LT precession rate was $37.2 \pm 7.2$ mas/yr. These measurements closely matched the predictions of GR, which were $6,606.1$ mas/yr for the geodetic effect and $39.2$ mas/yr for the LT precession \cite{GPBFinal}.

\vspace{2mm}

Beyond these classical treatments, more systematic and covariant formalisms have since been developed for describing gyroscopic spin precession. One approach formulates \cite{RinPer1990} the precession of point-like gyroscopes in stationary, axisymmetric spacetimes by analyzing their motion relative to a reference congruence of timelike worldlines. Another framework \cite{IyerVishveshwara} employs the Frenet–Serret description for quasi-Killing trajectories, emphasizing the role of congruence vorticity in governing spin evolution across diverse spacetimes. Nevertheless, understanding gyroscopic precession is not only crucial for testing GR but also for addressing a fundamental issue in gravitation theory: distinguishing black holes (BHs) from naked singularities (NaSs) \cite{CCDKNS}. A central question in this context is whether NaSs, if they do exist as end states of gravitational collapse, can be observationally distinguished from BHs. This distinction is crucial for making predictions about strong gravity regions in the universe, which are currently being probed by major astrophysical missions\cite{eventhorizontelescope}. In this context, Ref.\cite{CKJ} explored the possibility of distinguishing a Kerr black hole from a naked singularity using the spin precession of a test gyroscope attached to a static observer, though the analysis was restricted to regions outside the outer ergoregion (henceforth, simply the ergoregion). However, a recent study \cite{CCDKNS} has demonstrated that the divergence of the spin precession frequency at the ergoregion, previously reported for static observers in Ref. \cite{CKJ}, can be circumvented if the test gyroscope possesses a nonzero angular velocity $\Omega$ (referred to as stationary observers). While the four-velocity of `static gyroscopes' on the ergosurface becomes null ($u \cdot u = 0$), `stationary gyroscopes'—with an azimuthal component in their four-velocity—maintain a timelike norm. This modification allows for a well-defined study of gyroscopic behavior inside the ergoregion.  More generally, Ref.\cite{CCDKNS} derives the most comprehensive spin precession equation, which holds for any stationary and axisymmetric spacetime. This formulation ensures that stationary rotating gyroscopes do not exhibit a divergence in their precession frequency at the ergoregion, enabling an exploration of gyro behavior both inside and outside this region. Consequently, by analyzing the spin precession of a test gyroscope attached to a stationary observer, one may distinguish between a black hole and a naked singularity based on the gyro’s behavior.

\vspace{2mm}

In this article, we explore the precession of a gyroscope in the vicinity of the magnetized Kerr spacetime. We analyze how the magnetic field affects the spin precession of a gyro, and how the gyroscopic behavior can be used as probe to distinguish between these two compact objects. The inclusion of external magnetic field in a rotating spacetime introduces a unique tug-of-war phenomenon between the magnetic field and the ``gravitoelectromagnetic'' (GEM) effect \cite{CCDKNS}, leading to interesting features in spin precession. To derive the exact spin precession equation, we consider the Ernst \cite{ernstBHmag} and/or Wald \cite{WaldBHmag} solutions for magnetized Kerr spacetime, as these represent exact electrovacuum solutions of the Einstein-Maxwell equations following \cite{CCFaraday}. However, the Ernst solution \cite{CCFaraday} had certain drawbacks, which were addressed in \cite{Wild} to obtain a physically meaningful solution \cite{AG2}. This improved formulation has since been applied to study magnetic precession in BH systems with magnetized accretion disks \cite{AG3}, magnetic Penrose process \cite{CCPenrose}, and other related phenomena. Although gravitational energy generally dominates over electromagnetic energy, the two become comparable if the strength of the magnetic field ($B$) around a collapsed object of mass $M$ is of the order \cite{AG2, GaltsovBHMag}  

\begin{align}
 B \simeq B_{\rm max} \sim 2.4 \times 10^{19} \frac{M_{\odot}}{M} ~{\rm Gauss},
 \label{bmax}
\end{align}
where $M_{\odot}$ is the solar mass. While the magnetic field strength around a BH is typically much smaller than $B_{\rm max}$ (i.e., $B \ll B_{\rm max}$), studies suggest that for $B \sim B_{\rm max}$, the surrounding spacetime could be significantly distorted ~\cite{GaltsovBHMag,sha}. This highlights the importance of considering the magnetic field as a background field for testing the geometry around a collapsed object \cite{sha}. For this very reason, we do not impose any assumptions or approximations on the intensity of the magnetic field, thus allowing for a more general and accurate formulation of spin precession.

\vspace{2mm}

The structure of the paper is as follows. In Sec. \ref{sec2}, we introduce the Kerr spacetime immersed in a uniform magnetic field and discuss its fundamental properties. In Sec. \ref{sec3}, we summarize the derivation of the spin precession frequency for a test gyroscope in a general stationary and axisymmetric spacetime and apply this formalism to magnetized Kerr spacetime. In Sec. \ref{sec4}, we examine the characteristic features of spin precession in magnetized Kerr spacetime, highlighting the distinctions between black holes and naked singularities. The competing influences of the gravitomagnetic field and the external magnetic field on spin precession are analyzed in Sec. \ref{sec5}. In Sec. \ref{sec6}, we explore the effects of an external magnetic field on spin precession in the weak-field regime. The Lense-Thirring precession frequency in magnetized Kerr spacetime is derived and discussed in Sec. \ref{sec7}. In Sec. \ref{geo} we obtain the exact expression of geodetic precession in the non rotating magnetic Schwarzschild spacetime. Finally, in Sec. \ref{sec8}, we summarize our key findings and discuss the applications as well as limitations of our formulation.

\section{\label{sec2}Kerr spacetime immersed in a uniform magnetic field}

A Kerr black hole immersed in an external uniform magnetic field provides an exact electrovacuum solution to the Einstein-Maxwell equations. This spacetime retains its axial and time-translation symmetries, while the presence of a magnetic field modifies its structure, introducing additional effects due to the coupling between gravity and external magnetic field. The metric governing such a system in Boyer-Lindquist coordinates is given by \cite{AG1,AG2,CCPenrose}

\begin{align}
    ds^2 = \left( -\frac{\Delta}{A} dt^2 + \frac{dr^2}{\Delta} + d\theta^2 \right) \Sigma |\Lambda|^2 + \frac{A \sin^2\theta}{\Sigma|\Lambda|^2} (|\Lambda_0|^2d\phi - \varpi dt)^2. \label{kerrmagm}
\end{align}

Here, the functions defining the metric components are

\begin{align}
    \Delta &= r^2 + a^2 - 2Mr, & \quad \Sigma &= r^2 + a^2 \cos^2 \theta, \\
    A &= (r^2 + a^2)^2 - \Delta a^2 \sin^2 \theta, & \quad \varpi& = \frac{\alpha - \beta \Delta}{r^2 + a^2} + \frac{3}{4}a M^2 B^4. \label{varpi}
\end{align}

In the expression of $\varpi$

\begin{align}
    \alpha &= a(1-a^2M^2B^4), \\
   \beta &= \frac{a\Sigma}{A} + \frac{aMB^4}{16} \Big(-8r\cos^2\theta (3-\cos^2\theta) - 6 r\sin^4\theta + \frac{2a^2\sin^6\theta}{A} [2Ma^2 + r(a^2+r^2)] \nonumber \\ &\quad + \frac{4Ma^2\cos^2\theta}{A} [(r^2+a^2)(3-\cos^2\theta)^2 - 4a^2\sin^2\theta]  \Big).
\end{align}

Here, $M$ represents the mass, and $a$ denotes the spin parameter of the Kerr spacetime. An asymptotically uniform magnetic field $B$ is directed
along the polar axis, i.e., the vertical or
$z$-axis \cite{AG1}. The function $\Lambda(r,\theta)$ is a complex scalar modifying the geometry due to the presence of the magnetic field, expressed as

\begin{align}
     \Lambda  &\equiv \Lambda (r,\theta) = \text{Re} \ \Lambda + i \ \text{Im} \ \Lambda \nonumber \\ 
     & = 1 + \frac{B^2 \sin^2\theta}{4} \left [(r^2 +a^2) + \frac{2a^2Mr\sin^2\theta}{\Sigma} \right ] - i \frac{aB^2 M \cos\theta}{2} \left (3 - \cos^2\theta + \frac{a^2 \sin^4\theta}{\Sigma} \right ).
\end{align}

A crucial aspect of the metric structure is the presence of the Harrison-Ernst function $\Lambda(r,\theta)$ just before $d\phi$ in Eq. \eqref{kerrmagm}, which ensures the consistency of the solution with the Einstein-Maxwell equations. In particular, the term $|\Lambda_0|^2$ is introduced in the metric to remove the conical singularities on the polar axis ($\theta=0$) \cite{William1981,AG1,AG2}. This correction prevents the Ricci tensor from developing singularities along the polar axis and maintains the physical validity of the metric. The introduction of an external magnetic field leads to an azimuthal coordinate rescaling. This modification addresses conical singularities arising along the polar axis, requiring the azimuthal range to be adjusted from $2\pi$ to $2\pi |\Lambda_0|^2$ \cite{AG1, AG2}, where

\begin{align}
    |\Lambda_0|^2 = |\Lambda(r,0)|^2 = 1 + a^2 M^2 B^4.
\end{align}

Without this factor, the magnetized Kerr metric would no longer be a valid solution to the Einstein-Maxwell equations outside the polar axis, as it would cause the Ricci tensor to become singular at the polar axis. Despite the introduction of an external field, the event horizon structure remains unaltered. However, the presence of a magnetic field influences the location of the ergoregion (see \cite{Gibbons2013} for further details). Moreover, since our magnetized Kerr metric is not asymptotically flat rather, it has a nonvanishing magnetic field at infinity the curvature of spacetime is gradually dominated by the magnetic field as one moves away from the black hole. At large distances, the spacetime geometry approaches that of the Melvin universe \cite{Melvin}.

\section{\label{sec3}Spin Precession of a Test Gyroscope: Formalism and Application to Magnetized Kerr Spacetime}

Our goal is to study the spin precession of a test gyroscope in the magnetized Kerr spacetime. To achieve this, we consider stationary observers maintaining fixed radial ($r$) and polar ($\theta$) coordinates while rotating in a prograde direction around the central object. The four-velocity of such observers is given by \cite{CCDKNS}:

\begin{align} 
u^{\mu} = u^{t} (1,0,0,\Omega). 
\label{e9}
\end{align}
Here $t$ represents the time coordinate, and $\Omega$ is the angular velocity of the observer. Since the observers must follow timelike trajectories, the values of $\Omega$ are constrained accordingly. Throughout this work, we adopt stationary observers as the reference frame for analyzing gyroscope spin precession.

\vspace{2mm}

The formalism governing spin precession in a stationary and axisymmetric spacetime was developed in \cite{CCDKNS}. In this approach, a test gyroscope attached to a stationary observer moves along a Killing trajectory associated with a timelike Killing vector field $K$. The spin vector of the gyroscope undergoes Fermi-Walker transport along the observer’s four-velocity, and, the spin precession is measured relative to a locally constructed orthonormal triad \cite{StraumannGR}. This represents what a local observer would identify as non-rotating axes or ``axes at rest" \cite{StraumannGR}.  The spin precession is stuided here relative to this frame, which can also be interpreted as the ``Copernican system" \cite{StraumannGR, CCDKNS}.

Unlike the Kerr spacetime, the spin precession in the magnetized Kerr spacetime cannot be measured relative to a “distant star” \cite{GPBFinal}. As discussed in Sec. \ref{sec2}, the magnetized Kerr metric is not asymptotically flat, and therefore the concept of distant star serving as an inertial reference becomes ill-defined. In such scenarios, the adoption of the “Copernican system” may offer a more suitable framework. While any precession relative to the comoving vierbeins would typically correspond to precession relative to the distant stars (due to the Lorentz boost structure), in non-asymptotically flat spacetimes this correspondence fails.
In this work, we consider a stationary axisymmetric observer moving on a circular orbit in the magnetized Kerr spacetime with four-velocity \( u^\mu \). To analyze spin precession, we construct a physically motivated orthonormal triad \( \{ e^\mu_{(i)} \} \), defined in the local rest frame of the observer. The precession of the spin is then computed relative to this triad via Fermi-Walker transport. The construction leverages the spacetime's symmetries and background magnetic field and proceeds as follows:
\begin{itemize}
    \item The {\it first leg} of the triad, \( e^\mu_{(1)} \), is aligned with the direction of the external magnetic field, which in the magnetized Kerr spacetime is uniform and oriented along the axis of symmetry, typically the $z$-axis. This provides a global, physically measurable direction, identifiable by the observer using a local magnetic compass. Operationally, this leg is constructed as a spacelike unit vector from the electromagnetic field tensor \( F^{\mu\nu} \) via
\[
e^\mu_{(1)} = \frac{^*F^{\mu\nu} u_\nu}{\sqrt{(^*F^{\lambda\sigma} u_\sigma) (^*F_{\lambda\gamma} u^\gamma)}}
\]
where \( {^*}F^{\mu\nu} \) is the dual electromagnetic field tensor.

    \item The {\it second leg} of the triad, \( e^\mu_{(2)} \), is constructed by projecting the axial Killing vector \( \xi^\mu_{(\phi)} \) (associated with azimuthal symmetry) into the local rest space of the observer with four-velocity \( u^\mu \). This direction lies in the azimuthal plane, reflecting the axial symmetry of the spacetime and the symmetry of the circular orbit. The projection and normalization yield
\[
\bar{e}^\mu_{(2)} = \xi^\mu_{(\phi)} + \left( - u_\nu \xi^\nu_{(\phi)} \right) u^\mu, \quad
e^\mu_{(2)} = \frac{\bar{e}^\mu_{(2)}}{\sqrt{g_{\lambda\sigma} \bar{e}^\lambda_{(2)} \bar{e}^\sigma_{(2)}}}.
\]
This defines a spacelike unit vector orthogonal to both \( u^\mu \) and \( e^\mu_{(1)} \), with a clear operational meaning derived from the spacetime's axial symmetry.

    \item The {\it third leg} of the triad, \( e^\mu_{(3)} \), is constructed to complete the orthonormal frame using the Gram--Schmidt procedure. It lies in the radial direction as perceived by the observer, being orthogonal to both the magnetic field direction and the projected azimuthal direction. This leg is given by
\[
e^\mu_{(3)} = \frac{\epsilon^\mu_{\ \nu\rho\sigma} \, u^\nu \, e^\rho_{(1)} \, e^\sigma_{(2)}}{
\sqrt{g_{\kappa\zeta} \left( \epsilon^\kappa_{\ \lambda\gamma\delta} u^\lambda e^\gamma_{(1)} e^\delta_{(2)} \right)
\left( \epsilon^\zeta_{\ \eta\chi\psi} u^\eta e^\chi_{(1)} e^\psi_{(2)} \right)}}
\]
ensuring orthogonality of the triad within the observer’s local rest space. $\epsilon^\mu_{\ \nu\rho\sigma}$ is the Levi-Civita symbol.

\end{itemize}
This triad is not a coordinate artifact but is grounded in the physical symmetries and background field structure of the spacetime. It enables a meaningful, local, and covariant definition of spin precession in a setting where no global inertial frame exists. In the idealized magnetized Kerr setup--modeling a neutral mass in a uniform magnetic field--the magnetic field direction plays the role of a fixed, physically observable reference, analogous to the distant stars in asymptotically flat scenarios  \cite{personal, saadati}. Our construction is consistent with earlier frameworks \cite{RinPer1990,Jantzen1992} and provides a nontrivial and operationally relevant approach to studying precession in non-asymptotically flat backgrounds.
\vspace{.2cm}

In such a framework, the corresponding spin precession frequency $\Omega_p$ is determined by \cite{StraumannGR}
\begin{align}
\tilde{\Omega}_p &= \frac{1}{2K^2} *(\tilde{K} \wedge d\tilde{K}),
\label{GS}
\end{align}
where $ * $ symbolizes the Hodge dual, and $ \wedge $ indicates the wedge product. Additionally, $ \tilde{K} $ and $ \tilde{\Omega}_p $ refer to the one-forms associated with $ K $ and $ \Omega_p $, respectively. By employing the general timelike Killing vector expressed as $ K = \partial_0 + \Omega \partial_{\phi} $, one can derive the precession frequency applicable to any stationary and axisymmetric spacetime \cite{CCDKNS}:
\begin{align}
    \vec{\Omega}_p &= \frac{1}{2\sqrt{-g} \left(1 + 2\Omega \frac{g_{0\phi}}{g_{00}} + \Omega^2 \frac{g_{\phi\phi}}{g_{00}}  \right)} \nonumber \\
   &\left[-\sqrt{g_{rr}} \left[ \left(g_{0\phi,\theta} - \frac{g_{0\phi}}{g_{00}} g_{00,\theta} \right) + \Omega  \left(g_{\phi\phi,\theta} - \frac{g_{\phi\phi}}{g_{00}} g_{00,\theta} \right) + \Omega^2  \left(\frac{g_{0\phi}}{g_{00}} g_{\phi\phi,\theta} - \frac{g_{\phi\phi}}{g_{00}} g_{0\phi,\theta} \right) \right] \hat{r} \right. \nonumber \\
   &\quad \left.+ \sqrt{g_{\theta\theta}} \left[ \left(g_{0\phi,r} - \frac{g_{0\phi}}{g_{00}} g_{00,r} \right) + \Omega  \left(g_{\phi\phi,r} - \frac{g_{\phi\phi}}{g_{00}} g_{00,r} \right) + \Omega^2  \left(\frac{g_{0\phi}}{g_{00}} g_{\phi\phi,r} - \frac{g_{\phi\phi}}{g_{00}} g_{0\phi,r} \right) \right] \hat{\theta} \right] \label{Omegavecp}.
\end{align}

For a more detailed derivation of Eq. \eqref{Omegavecp}, we refer the reader to Ref. \cite{CCDKNS}. We now apply the formalism developed above to the magnetized Kerr spacetime to analyze the behavior of a test gyroscope both inside and outside the ergoregion. The metric components can be directly read from Eq. \eqref{kerrmagm}, and using that we obtain

\begin{align}
     \sqrt{-g} = \left(1 + a^2 B^4 M^2\right) |\Lambda|^2 \ \Sigma \sin\theta.
\end{align}
By substituting the metric components of the magnetized Kerr spacetime into Eq.~\eqref{Omegavecp}, we obtain the spin precession frequency of a gyroscope as

\begin{align}
    \vec{\Omega}_p \approx \frac{\mathcal{U}\sqrt{\Delta}\cos\theta \hat{r} + \mathcal{V} \sin\theta \hat{\theta}}{\Sigma^{3/2} \Big[(\Sigma - 2Mr)+ 4\Omega M a r \sin^2 \theta - \Omega^2 \sin^2 \theta[\Sigma(r^2+a^2) + 2Ma^2r\sin^2\theta]\Big]} + B^2 \vec{\mathcal{K}} (r,\theta, M, a, \Omega) + \mathcal{O}(B^3), \label{Spinmagkerr}
\end{align}
where,
\begin{align}
    \mathcal{U} &= 2 aMr - \frac{\Omega}{8} [8r^4+8a^2r^2+16a^2Mr+3a^4+4a^2(2\Delta-a^2)\cos2\theta+a^4\cos4\theta] + 2\Omega^2 a^3Mr\sin^4\theta, \nonumber \\ 
    \mathcal{V} &= aM(r^2-a^2\cos^2\theta) + \Omega\Big[a^4r\cos^4\theta + r^2(r^3-3Mr^2-a^2M(1+\sin^2\theta)) \nonumber \\ & \quad +a^2\cos^2\theta(2r^3-Mr^2+a^2M(1+\sin^2\theta))\Big] + \Omega^2 aM\sin^2\theta[r^2(3r^2+a^2)+a^2\cos^2\theta(r^2-a^2)].
    \label{UV}
\end{align}

In this context, $\vec{\mathcal{K}}$ captures the leading-order contribution of the magnetic field to spin precession. Although we have obtained exact analytical expressions for $\vec{\Omega}_p$ in magnetized Kerr spacetime, we omit their explicit presentation due to their lengthy and cumbersome form. Instead, we employ graphical analysis to convey their physical meaning more intuitively. The complete analytical results are available upon request. All figures in this work are generated using the exact form of $\vec{\Omega}_p$. Notably, in Eq. \eqref{Spinmagkerr}, the first term arises from the Kerr spacetime and matches the results obtained in \cite{CCDKNS}. In the above expression for timelike observers, $\Omega$ has the restriction are confined within the bounds given by

\begin{align}
    \Omega_-(r, \theta) < \Omega(r, \theta) < \Omega_+(r, \theta),
\end{align}
where \footnotetext{Once again, we present the analytical expression up to $\mathcal{O}(B^2)$ to illustrate the dependence of observables on the magnetic field. However, all plots are generated using the exact expressions, which are omitted here due to their length but are available upon request.},
\begin{align}
    \Omega_{\pm} \approx \frac{2Mar\sin{\theta} \pm \Sigma \sqrt{\Delta}}{\sin{\theta} [\Sigma(r^2 + a^2) + 2 Ma^2r\sin^2{\theta}]} \pm \frac{1}{2} \sqrt{\Delta} \sin{\theta} B^2 + \mathcal{O}(B^3). \label{rangemagkerr}
     \footnotemark
\end{align} 

Furthermore, we stress that neither the event horizon nor the ring singularity ($r = 0, \theta = \pi / 2$) admits well-defined solutions for $\Omega$ due to the imposed inequality. This implies the absence of stationary timelike observers at these locations. As a result, the spin precession framework developed here does not hold in these extreme cases.

\vspace{2mm}

\begin{figure}[h!]
\begin{center}
\subfigure[$a_*=0.9$, $B=0.1 M^{-1}$]{
\includegraphics[width=3in,angle=0]{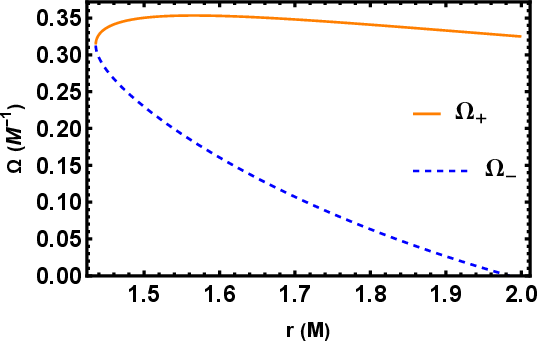}}
\subfigure[$a_*=2$, $B=0.1 M^{-1}$]{
\includegraphics[width=3in,angle=0]{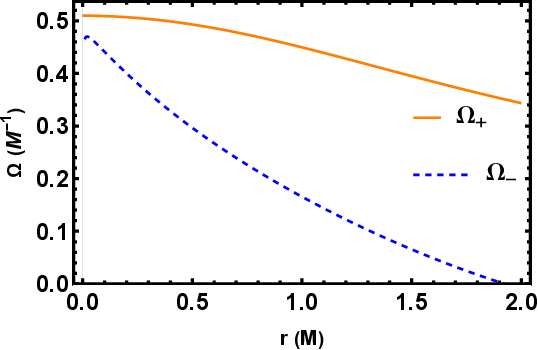}}
\caption{\label{Omega_pm}The frequency of a stationary gyroscope, $\Omega$, can take values in the range $\Omega_{-} \leq \Omega \leq \Omega_{+}$ at any $(r, \theta)$ and they represent the lower and upper limits of the frequency, respectively. In these plots, we have taken $B = 0.1 M^{-1}$. For BH, we can see from panel (a) that $\Omega_{\pm}$ meet at the horizon. For NaS, the two curves remains distinct in the limit $r \to 0$ which is illustrated in panel (b). See Sec. \ref{sec3} for details.}
\end{center}
\end{figure}

Panel (a) of Fig. \ref{Omega_pm} illustrates that for a BH with $a_* = 0.9$ in external magnetic field strength of $B = 0.1 M^{-1}$, the frequencies $\Omega_+|_{\theta=\frac{\pi}{2}}$ and $\Omega_-|_{\theta=\frac{\pi}{2}}$ coincide at event horizon $r \to r_+$. In contrast, for NaS, the behavior of $\Omega_{\pm}$ differs significantly. As shown in panel (b) of Fig. \ref{Omega_pm} for a NaS with $a_* = 2$ and the same magnetic field strength, the two curves do not intersect in the limit $r \to 0$, highlighting a key distinction from the BH case.

\vspace{1mm}

To examine the behavior of the gyroscope in the vicinity of a BH and NaS, we can now plot $\Omega_p = |\vec{\Omega}_p|$ as a function of $r$ at different values of $\theta$ for observers with varying $\Omega$. To make this comparison easier, we introduce the parameter $q$ to explore the range of permitted values for $\Omega$ as follows:

\begin{align}
    \Omega &= q\Omega_+ + (1-q) \Omega_- \nonumber \\
    &= \omega - (1-2q) \sqrt{\omega^2 - \frac{g_{tt}}{g_{\phi\phi}}} \nonumber \\
    &\approx \frac{2Mar\sin{\theta}-(1 -2q) \Sigma \sqrt{\Delta}}{\sin{\theta}[\Sigma(a^2 + r^2) + 2Ma^2r\sin^2{\theta}]} - \frac{1}{2} (1-2q)\sqrt{\Delta}\sin{\theta} B^2 + \mathcal{O}(B^3)  \label{qomega}
\end{align}

where $ 0 < q < 1 $, and $\omega = -g_{t\phi}/g_{\phi\phi}$ . It is evident that the parameter $ q $ spans the full range of $ \Omega $, extending from $ \Omega_+ $ to $ \Omega_- $.

\vspace{2mm}

A particularly significant case arises for $q = 0.5$, where

\begin{align}
    \Omega \big|_{q=0.5} = \omega &= \frac{1}{1+a^2 B^4 M^2} \left( \frac{3}{4} a B^4M^2  + \frac{\alpha-\beta \Delta}{a^2+r^2}\right) \nonumber \\ 
    &\approx \frac{2Mar}{(r^2 + a^2)^2-a^2\Delta\sin^2\theta} + \mathcal{O}(B^3)
    \label{ZAMO1}
\end{align}

Observers with this angular velocity are known as zero-angular-momentum observers (ZAMOs). In this scenario, test gyroscopes attached to the stationary observers are considered non-rotating with respect to the local spacetime, meaning the angular momentum of a locally non-rotating observer is zero. The concept of ZAMOs was first introduced by Bardeen \cite{Bardeen1970} (see also \cite{Bardeen1972}). As demonstrated in Ref.\cite{CCDKNS}, the precession frequency of a gyroscope attached to a ZAMO in the standard Kerr spacetime exhibits behavior distinct from that of gyroscopes attached to other observers with different angular velocities. Thus, it is of particular interest to investigate how the gyroscope’s precession frequency for ZAMOs behaves in the presence of external magnetic field. 

\vspace{2mm}

\section{\label{sec4}Spin Precession Signatures in Magnetized Kerr Spacetimes: Black Holes Versus Naked Singularities}

In this section, we highlight the distinct differences in the behavior of $\Omega_p$ for stationary gyroscopes in the presence of BHs and NaSs. We show that $\Omega_p$ grows arbitrarily large for gyroscopes arbitrarily close to the BH horizon $r \sim r_+$ for all values of $q$ except $q=0.5$ (corresponding to the ZAMO frequency). In contrast, for NaS, $\Omega_p$ remains finite everywhere except at $r=0$ and $\theta \sim \pi/2$ (near the ring singularity). We identify distinguishing features in the radial profiles of $\Omega_p$ for both BHs and NaS in the presence of a non-zero external magnetic field, which we discuss in detail below.

\begin{figure}[!h]
\begin{center}
\subfigure[~$q=0.1, \theta=10^0$]{
\includegraphics[width=2in,angle=0]{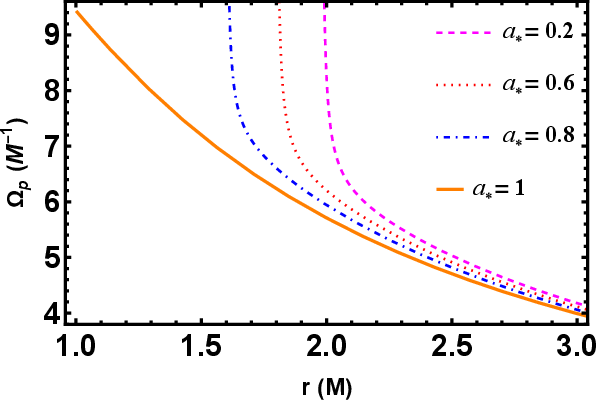}}
\subfigure[~$q=0.1, \theta=50^0$]{
\includegraphics[width=2in,angle=0]{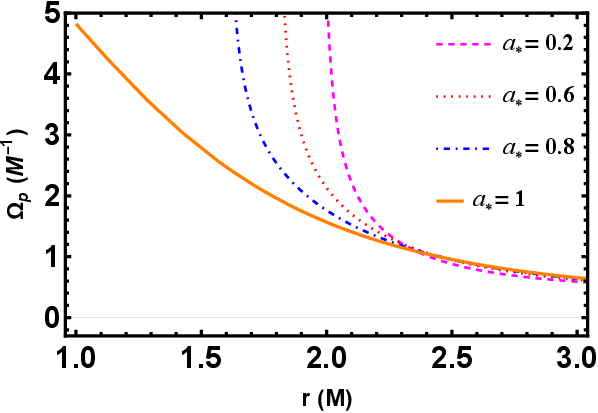}}
\subfigure[~$q=0.1, \theta=90^0$]{
\includegraphics[width=2in,angle=0]{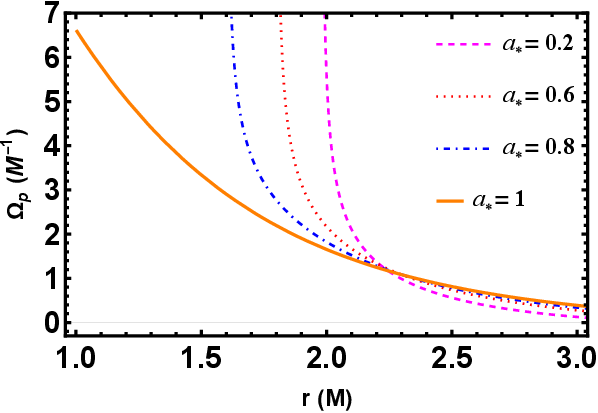}}
\subfigure[~$q=0.5, \theta=10^0$]{
\includegraphics[width=2in,angle=0]{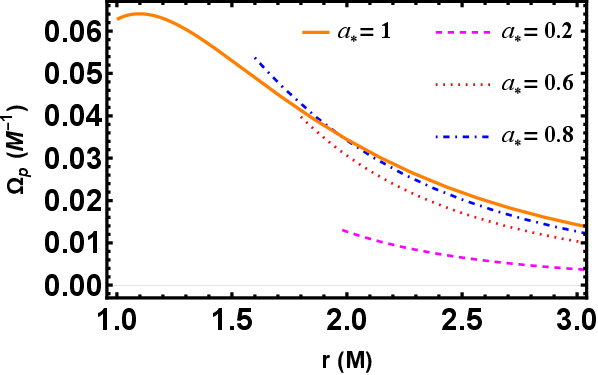}}
\subfigure[~$q=0.5, \theta=50^0$]{
\includegraphics[width=2in,angle=0]{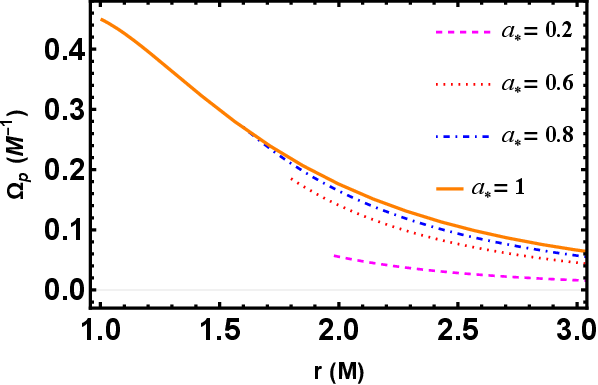}}
\subfigure[~$q=0.5, \theta=90^0$]{
\includegraphics[width=2in,angle=0]{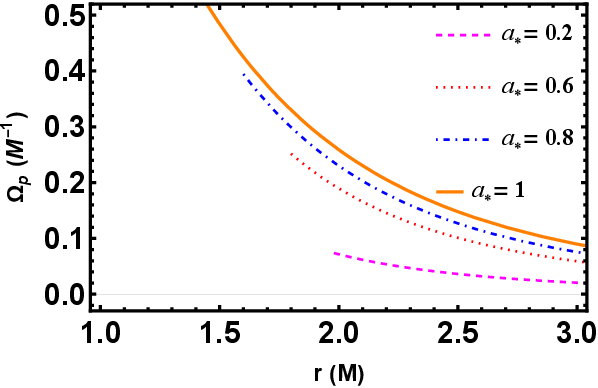}}
\subfigure[~$q=0.9, \theta=10^0$]{
\includegraphics[width=2in,angle=0]{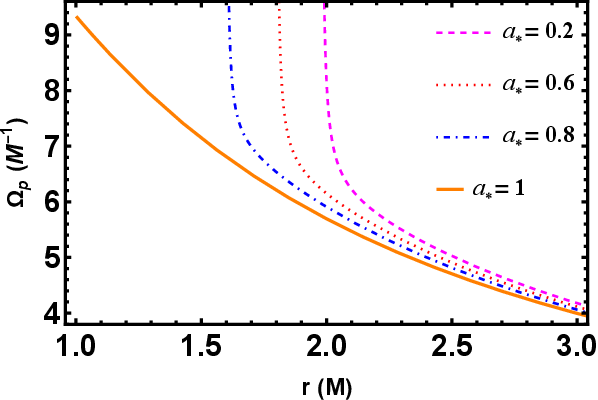}}
\subfigure[~$q=0.9, \theta=50^0$]{
\includegraphics[width=2in,angle=0]{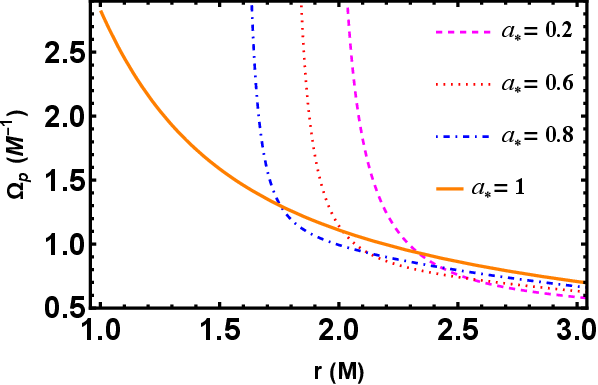}}
\subfigure[~$q=0.9, \theta=90^0$]{
\includegraphics[width=2in,angle=0]{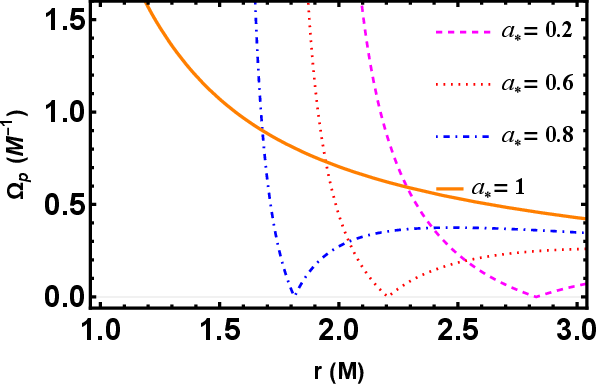}}
\caption{\label{BHGen} $\Omega_p$ (in $M^{-1}$) versus $r$ (in $M$) for $B = 0.1 M^{-1}$ is shown for various values of $a_*$, $q$, and $\theta$, ranging from the horizon radius ($r_+$) to $r = 3M$, to highlight the behavior of $\Omega_p$ in the strong gravity regime in the presence of strong magnetic fields. This figure clearly illustrates that $\Omega_p$ becomes arbitrarily large near the event horizon for all values of $a_*$, $q$, and $\theta$ with $q=0.5$ as exception. See Sec. \ref{sec4} for details.}
\end{center}
\end{figure}

\begin{figure}[!h]
\begin{center}
\subfigure[~$q=0.1, \theta=10^0$]{
\includegraphics[width=2in,angle=0]{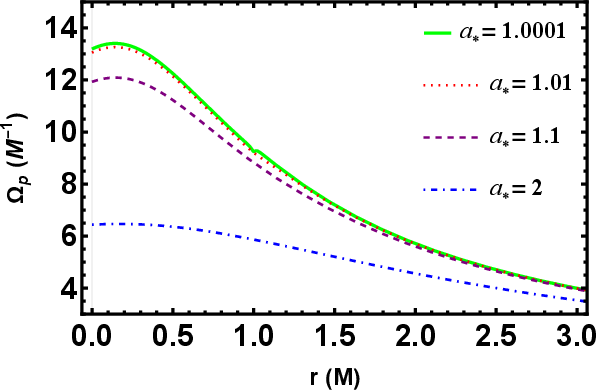}}
\subfigure[~$q=0.1, \theta=50^0$]{
\includegraphics[width=2in,angle=0]{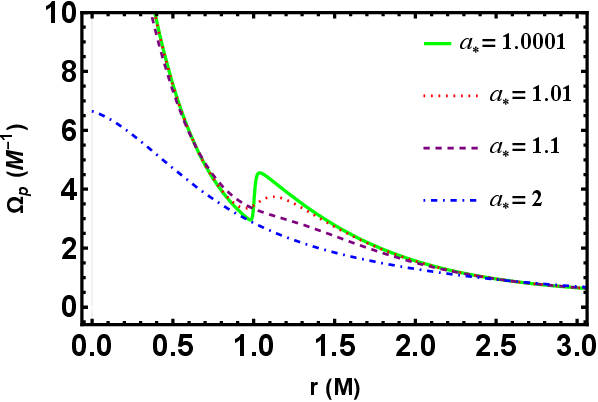}}
\subfigure[~$q=0.1, \theta=90^0$]{
\includegraphics[width=2in,angle=0]{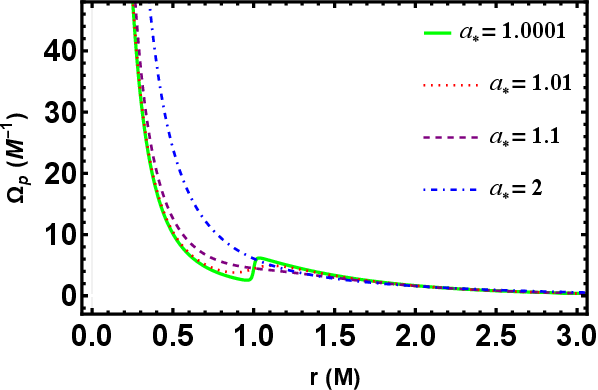}}
\subfigure[~$q=0.5, \theta=10^0$]{
\includegraphics[width=2in,angle=0]{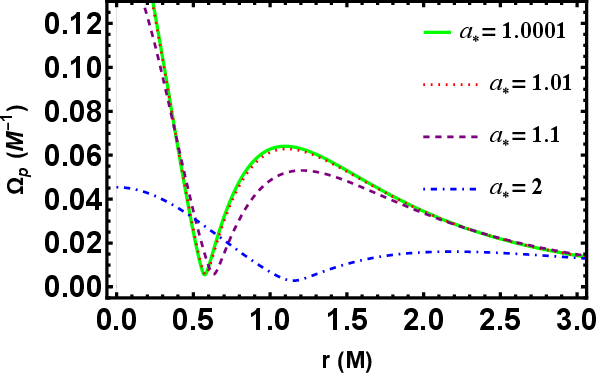}}
\subfigure[~$q=0.5, \theta=50^0$]{
\includegraphics[width=2in,angle=0]{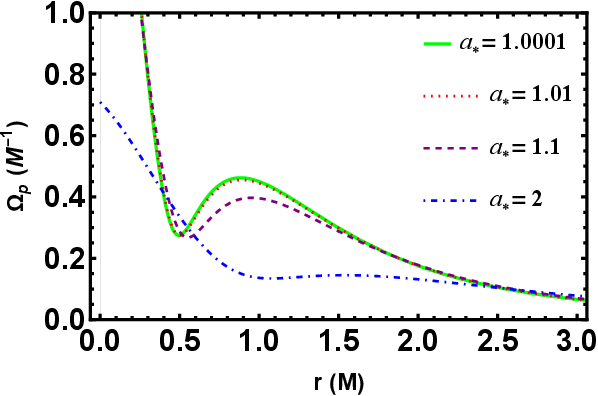}}
\subfigure[~$q=0.5, \theta=90^0$]{
\includegraphics[width=2in,angle=0]{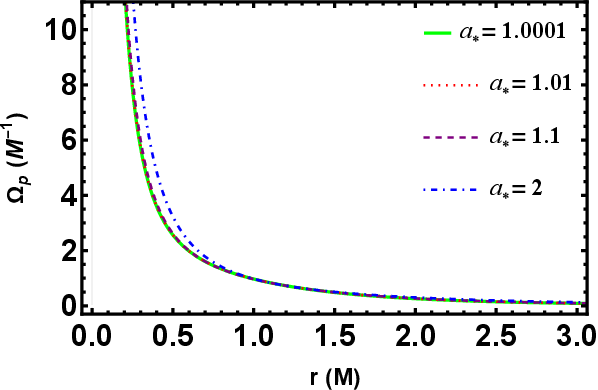}}
\subfigure[~$q=0.9, \theta=10^0$]{
\includegraphics[width=2in,angle=0]{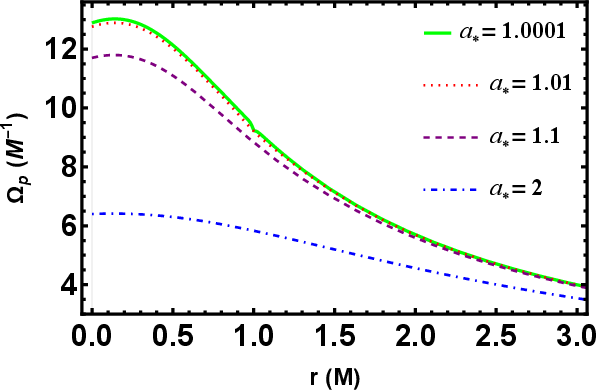}}
\subfigure[~$q=0.9, \theta=50^0$]{
\includegraphics[width=2in,angle=0]{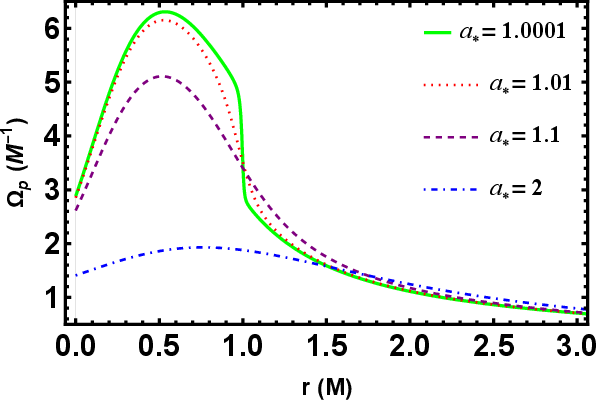}}
\subfigure[~$q=0.9, \theta=90^0$]{
\includegraphics[width=2in,angle=0]{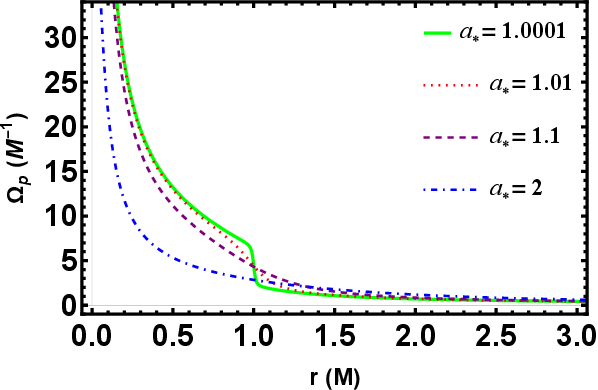}}
\caption{\label{NSGen} $\Omega_p$ (in $M^{-1}$) versus $r$ (in $M$) for $B = 0.1 M^{-1}$ is shown for various values of $a_*$, $q$, and $\theta$, ranging from $r=0$ to $r = 3M$, to highlight the behavior of $\Omega_p$ in the strong gravity regime in the presence of strong magnetic fields. This figure clearly illustrates that $\Omega_p$ is finite at $r \rightarrow 0$ for $\theta \neq \pi/2$, whereas it becomes arbitrarily large near the ring singularity for all values of $a_*$ and $q$. See Sec. \ref{sec4} for details.}
\end{center}
\end{figure}

\vspace{2mm}

To investigate the impact of an external magnetic field on the precession frequency first, we consider BHs immersed in an external magnetic field of strength $ B = 0.1 M^{-1}$. We analyze the spin precession frequency $\Omega_p$ by plotting its radial variation for different spin parameters: $a_* = 0.2, 0.6, 0.8, 1$ . Our results in Fig.~\ref{BHGen} show that $\Omega_p $ remains finite and smooth both inside and outside the ergoregion but becomes arbitrarily large near the event horizon for all values of $a_*, q$, and  $\theta$, except when $q=0.5$. The top row of Fig.~\ref{BHGen} shows that for $q < 0.5$, $\Omega_p$ decreases monotonically. However, for $q > 0.5$, local minima starts to emerge as $\theta$ increases, as seen in the panels of the last row of Fig.~\ref{BHGen}. Interestingly, for ZAMO observers ($\Omega = \omega$), the spin precession frequency remains finite even for gyroscopes orbiting arbitrarily close to the horizon ($r_h \equiv M(1+\sqrt{1-a_*^2})$), as depicted in the second row of Fig.~\ref{BHGen}. Analytically, this can be understood as follows: while the radial contribution vanishes as the test gyroscope approaches $r_h$, the angular contribution remains finite. This ensures a well-defined spin precession equation at $r = r_h$, where Eq.~\eqref{Spinmagkerr} simplifies to
\begin{align}
   \Omega_p\big|_{r \to r_h} \approx \frac{a M \sin \theta  \left(-a^4+a^2 \left(r_h^2-a^2\right) \cos 2 \theta +3 a^2 r_h^2+6 r_h^4\right)}{2 \left(a^2 \cos ^2\theta +r^2_h\right)^{3/2} \left(a^2+r_h^2\right)^2}+\mathcal{O}\left(B^3\right).
   \label{ZAMOfinite}
\end{align}
Thus, Eq.~\eqref{ZAMOfinite} clarifies why the spin precession frequency remains finite in the ZAMO case. 
\vspace{2mm}

On the other hand, for NaSs with spin parameters $a_* = 1.0001, 1.01, 1.1, 2$ immersed in an external magnetic field of strength $ B = 0.1 M^{-1}$, $\Omega_p$ exhibits a fundamentally different character, as illustrated in Fig.~\ref{NSGen}. The values of $a_*$ are selected at non-uniform intervals to probe potential structural variations, particularly for near-extremal NaSs. Unlike the BH case, the spin precession frequency $\Omega_p$ remains finite and well-behaved across the domain $0 < \theta \lesssim 90^\circ$, demonstrating that the absence of an event horizon significantly alters the nature of spin precession. However, as one approaches  $r \to 0$ along the equatorial plane $\theta = \pi/2$, $\Omega_p$ grows arbitrarily large due to the presence of the ring singularity, marking a stark contrast to the BH case, where it becomes arbitrarily large near the event horizon, far from $r = 0$. Furthermore, as seen in panels of Fig. \ref{NSGen}, for $q \geq 0.5$, the radial profile of $\Omega_p$ exhibits additional structure, with local minima and maxima appearing at certain angles. We also note here that near-extremal NaSs appear to have additional characteristic features that could be used to distinguish them from other generic higher-spin NaSs, as can be seen clearly from the panels of Fig. \ref{NSGen}.

\vspace{2mm}

In a similar approach to the experiment proposed in \cite{CCDKNS}, which distinguishes Kerr BHs from NaSs using gyroscope precession frequencies, we suggest that this method could also be applied to magnetized Kerr spacetimes. For gyroscopes attached to stationary observers moving along circular orbits at constant $r$ and $\theta$, the precession frequency $\Omega_p$ of the gyroscopes can be studied to distinguish between a BH and a NaS:

\begin{itemize}
  \item If the precession frequency $\Omega_p$ becomes arbitrarily large as the observer approaches the central object in \textit{all} directions, it indicates the presence of a black hole.
  \item If the precession frequency only becomes arbitrarily large in one direction (specifically, along the equatorial plane $\theta = \pi/2$), the object is a naked singularity.
\end{itemize}

For magnetized Kerr spacetimes, the same reasoning could be applied. The presence of a magnetic field might influence the gyroscope precession behavior, but the fundamental principle remains unchanged. If $\Omega_p$ diverges in all directions, it suggests a black hole, while divergence in only one direction, i.e., along the equatorial plane, could indicate a naked singularity.

\section{\label{sec5}Tug of War: Gravitoelectromagnetism Versus Magnetic Field in Spin Precession}

 To examine the effect of magnetic field on spin precession, we consider a gyroscope attached to a stationary observer moving along a circular orbit at fixed $(r, \theta)$ with a nonzero azimuthal velocity. The observer’s motion is constrained within an allowable range of angular velocities, which we express using the parameter $q$. Now, considering a BH with $a_*=0.9$ and a NaS with $a_*=1.1$, we analyze the behavior of $\Omega_p$, as a function of radial coordinate $r$, while varying the magnetic field strength $B$, for selected values of $\theta$ and $q$. By constructing these profiles, we can systematically assess the role of the magnetic field in modifying the spin precession dynamics.

\begin{figure}[!h]
\begin{center}
\subfigure[~$q=0.1, \theta=10^0$]{
\includegraphics[width=2in,angle=0]{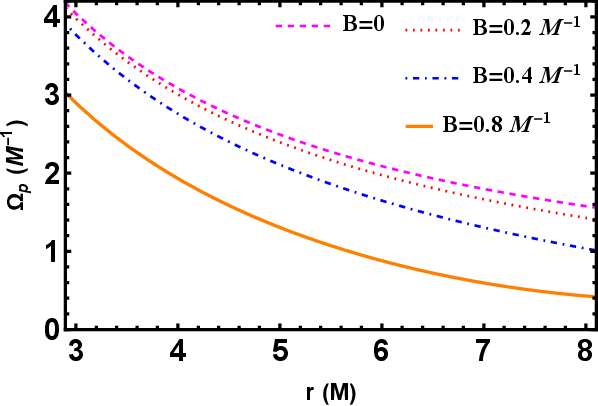}}
\subfigure[~$q=0.1, \theta=50^0$]{
\includegraphics[width=2in,angle=0]{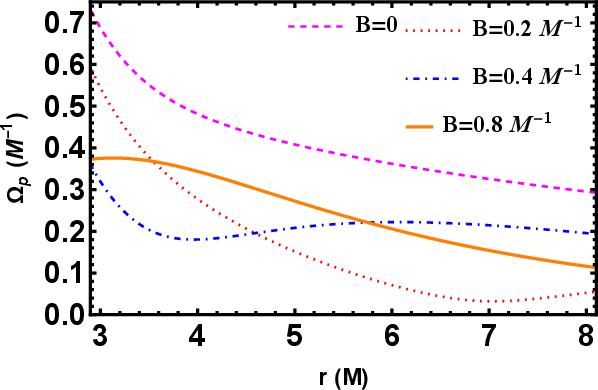}}
\subfigure[~$q=0.1, \theta=90^0$]{
\includegraphics[width=2in,angle=0]{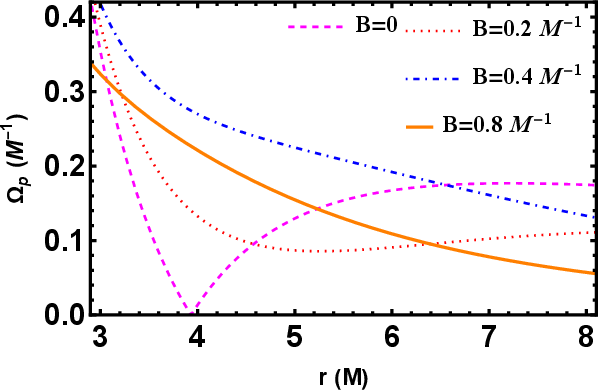}}
\subfigure[~$q=0.5, \theta=10^0$]{
\includegraphics[width=2in,angle=0]{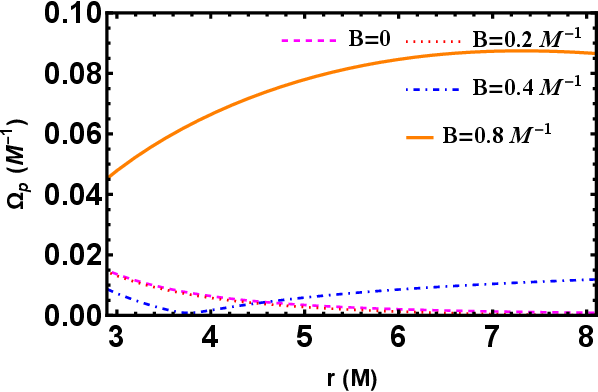}}
\subfigure[~$q=0.5, \theta=50^0$]{
\includegraphics[width=2in,angle=0]{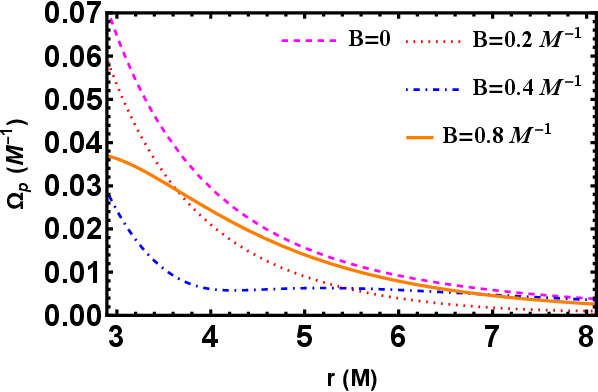}}
\subfigure[~$q=0.5, \theta=90^0$]{
\includegraphics[width=2in,angle=0]{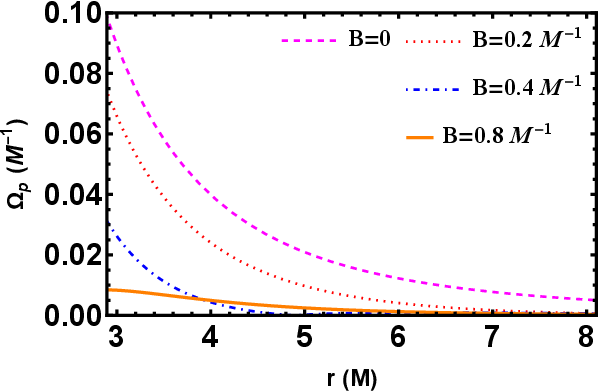}}
\subfigure[~$q=0.9, \theta=10^0$]{
\includegraphics[width=2in,angle=0]{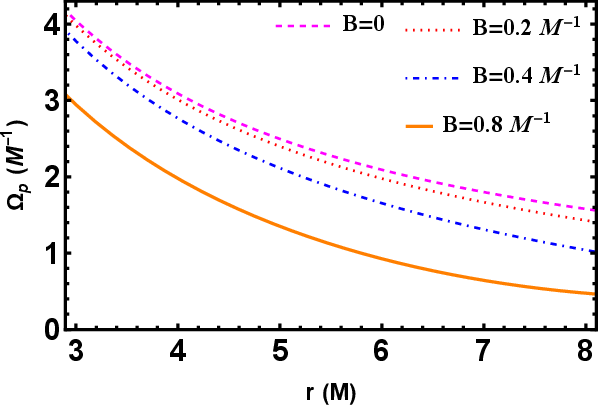}}
\subfigure[~$q=0.9, \theta=50^0$]{
\includegraphics[width=2in,angle=0]{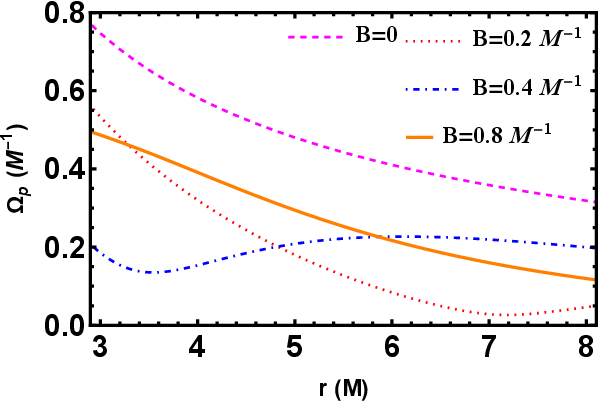}}
\subfigure[~$q=0.9, \theta=90^0$]{
\includegraphics[width=2in,angle=0]{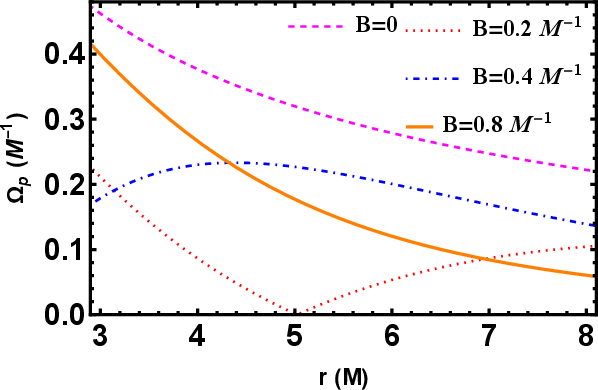}}
\vspace{2mm}
\caption{\label{BHTOW}$\Omega_p$ (in $M^{-1}$) versus $r$ (in $M$) for $a_*=0.9$ is shown for various values of $B$, $q$, and $\theta$, ranging from ranging from $3M$ to $8M$. The plots clearly illustrates the interplay between the GEM field and the external magnetic field, highlighting the tug-of-war between these two effects. See Sec. \ref{sec5} for details.}
\end{center}
\end{figure}

\begin{figure}[!h]
\begin{center}
\subfigure[~$q=0.1, \theta=10^0$]{
\includegraphics[width=2in,angle=0]{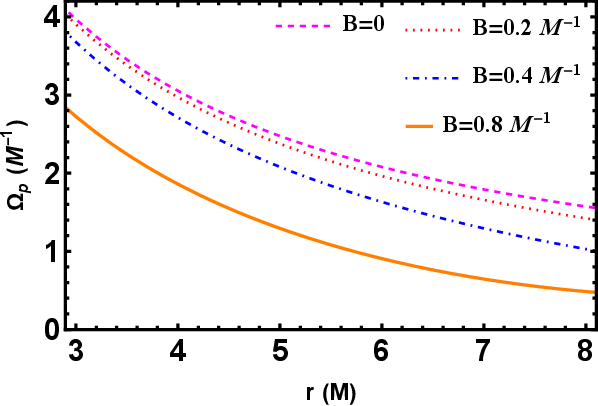}}
\subfigure[~$q=0.1, \theta=50^0$]{
\includegraphics[width=2in,angle=0]{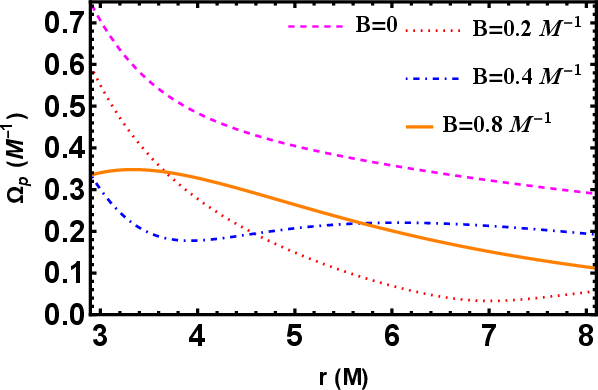}}
\subfigure[~$q=0.1, \theta=90^0$]{
\includegraphics[width=2in,angle=0]{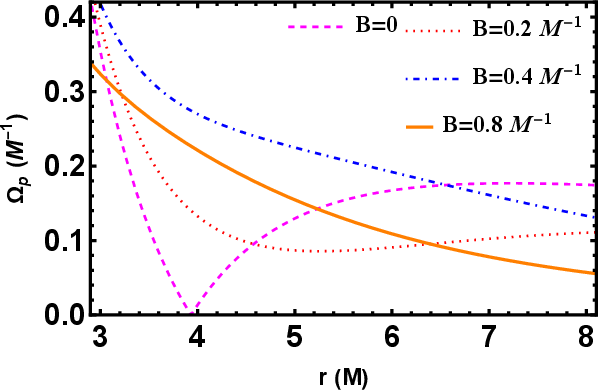}}
\subfigure[~$q=0.5, \theta=10^0$]{
\includegraphics[width=2in,angle=0]{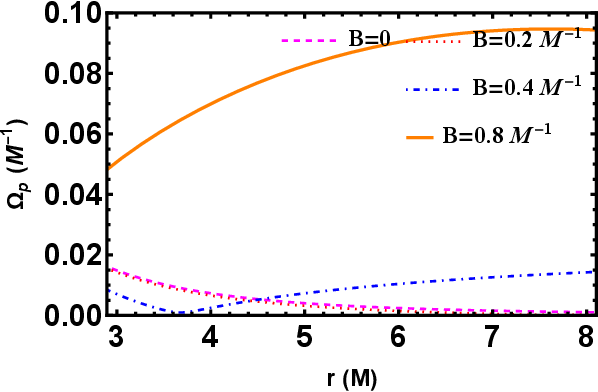}}
\subfigure[~$q=0.5, \theta=50^0$]{
\includegraphics[width=2in,angle=0]{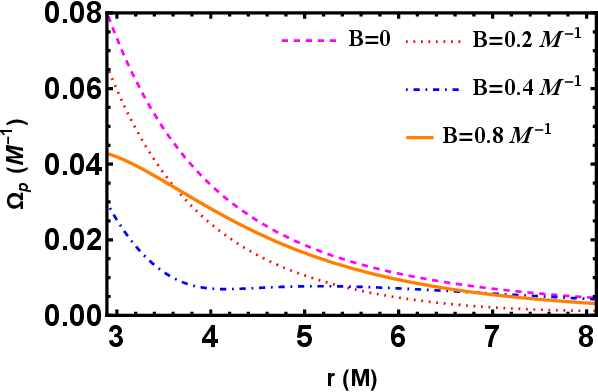}}
\subfigure[~$q=0.5, \theta=90^0$]{
\includegraphics[width=2in,angle=0]{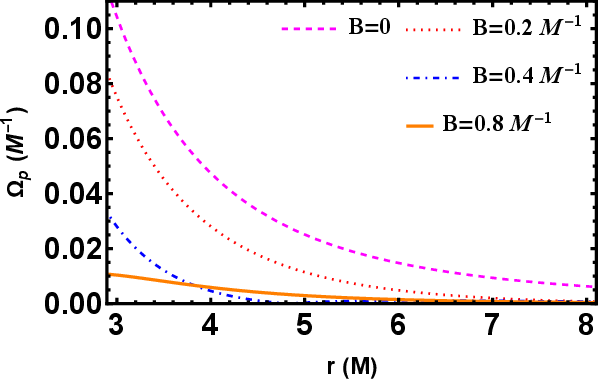}}
\subfigure[~$q=0.9, \theta=10^0$]{
\includegraphics[width=2in,angle=0]{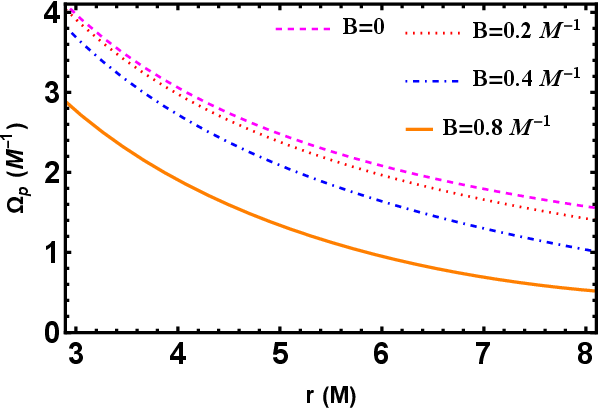}}
\subfigure[~$q=0.9, \theta=50^0$]{
\includegraphics[width=2in,angle=0]{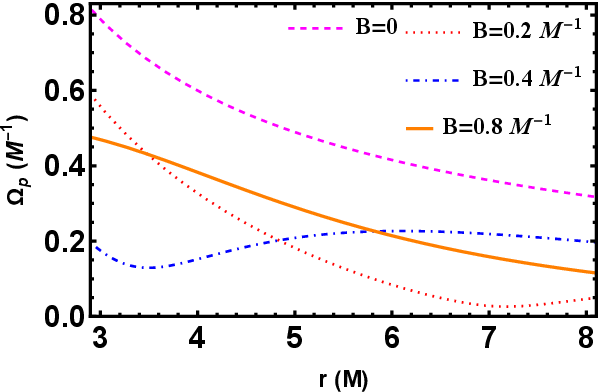}}
\subfigure[~$q=0.9, \theta=90^0$]{
\includegraphics[width=2in,angle=0]{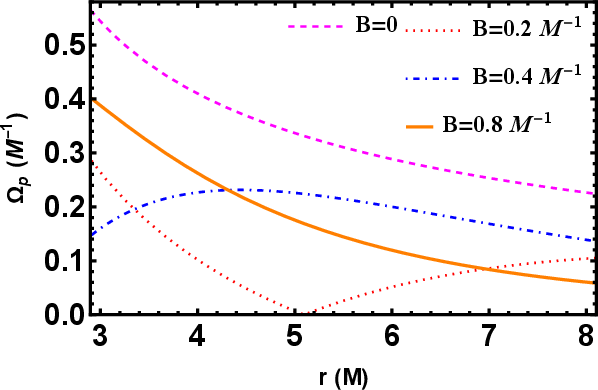}}
\caption{\label{NSTOW} $\Omega_p$ (in $M^{-1}$) versus $r$ (in $M$) for $a_*=1.1$ is shown for various values of $B$, $q$, and $\theta$, ranging from $3M$ to $8M$. The plots clearly illustrates the interplay between the GEM field and the external magnetic field, highlighting the tug-of-war between these two effects. See Sec. \ref{sec5} for details.}
\end{center}
\end{figure}

In Fig.~\ref{BHTOW}, we present the modulus of the spin precession frequency $\Omega_p$, as a function of the radial coordinate $r$ around a magnetized Kerr BH with a spin parameter of $a_* = 0.9$. To investigate the influence of the magnetic field on spin precession, we vary its strength while also including the $B = 0$ case as a reference for the standard Kerr scenario. The top row of Fig.~\ref{BHTOW}, panels (a), (b), and (c), corresponds to observers with $\Omega < \omega$ for $q = 0.1$. At lower values of $\theta$, $\Omega_p$ decreases monotonically. However, as $\theta$ increases from $10^\circ$ to $50^\circ$, we observe a more intricate interplay between different values of $B$, a behavior that persists even at $\theta = 90^\circ$. A similar trend is evident in the bottom row, panels (g), (h), and (i), for observers with $\Omega > \omega$ and $q = 0.9$. A similar trend can be noticed for the case of ZAMOs, shown in the middle row, panels (d), (e), and (f), where $\Omega = \omega$ with $q = 0.5$. This behavior arises due to a competing influence between two distinct effects: the GEM field of the BH (which is induced by its spin $a$, and mass $M$) and the externally imposed magnetic field. In essence, the magnetic field introduces an additional effect that modifies spin precession, sometimes amplifying and sometimes counteracting the effects of GEM. This competition leads to what can be thought of as a tug-of-war between the two influences.

\vspace{2mm}

Similar to the BH case, in Fig.~\ref{NSTOW}, we plot $\Omega_p$ as a function of the radial coordinate $r$ for a magnetized Kerr NaS with a spin parameter of $a_* = 1.1$. The results reveal that the NaS also exhibits the same underlying tug-of-war effect between the GEM field and the external magnetic field. This effect is evident in the first row of Fig.~\ref{NSTOW}, panels (a), (b), and (c), for observers with $\Omega < \omega$ ($q = 0.1$), as well as in the last row, panels (g), (h), and (i), for observers with $\Omega > \omega$ ($q = 0.9$). For the case of ZAMO ($\Omega = \omega$), depicted in the middle row panels (d), (e), and (f) with  $q = 0.5$, a similar trend is observed, closely resembling the behavior seen in the BH scenario illustrated in Fig.~\ref{BHTOW}.

 \vspace{2mm}
Interestingly, Figs.~\ref{BHTOW} and~\ref{NSTOW} reveal a striking similarity in the tug-of-war between BHs and NaS, though minute differences in precession frequencies emerge near the compact object. This suggests that the competition between the GEM field and external magnetic fields is largely—but not entirely—independent of the central object's nature. It is important to note that the tug-of-war is prominent at intermediate distances rather than close to the central objects. Near the compact object, the GEM field dominates,  suppressing the influence of the external magnetic field. However, as we move outward, the effect of the external magnetic field becomes more significant, allowing it to compete with the GEM field. This happens because of the increase in magnetic energy $\int_{\tau} B^2 d\tau$ (where, $d \tau$ is the volume element) at larger distances. This competition between the two fields—where neither completely dominates the other—gives rise to the observed tug-of-war effect in spin precession at intermediate distances. 

\section{\label{sec6}Spin Precession in the Weak-Field Regime: Effects of External Magnetic Field}

\begin{figure}[h!]
\begin{center}
\subfigure[$a_*=0.9$, $q=0.1$,$\theta=90^\circ$]{
\includegraphics[width=2in,angle=0]{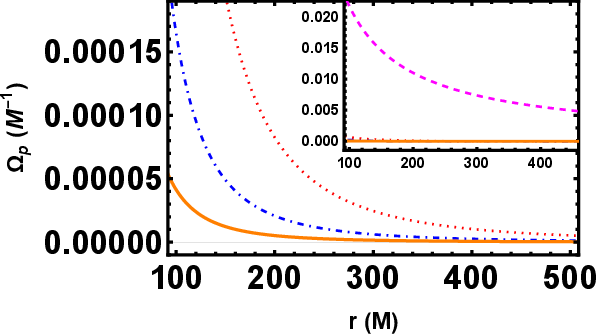}}
\subfigure[$a_*=0.9$, $q=0.5$,$\theta=90^\circ$]{
\includegraphics[width=2in,angle=0]{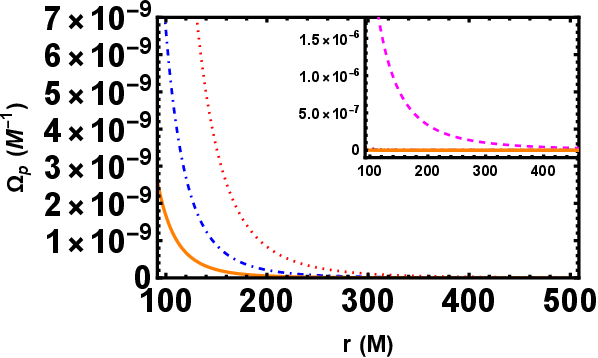}}
\subfigure[$a_*=0.9$, $q=0.9$,$\theta=90^\circ$]{
\includegraphics[width=2in,angle=0]{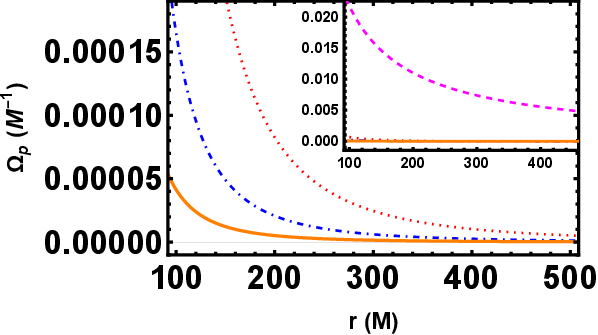}}
\includegraphics[width=3in,angle=0]{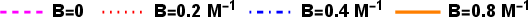}
\caption{\label{larger} $\Omega_p$ (in $M^{-1}$), is plotted as a function of $r$ (in $M$) in the weak-field regime ($r \gg M$) for a magnetized Kerr BH. Panels (a), (b), and (c) correspond to $q = 0.1$, $q = 0.5$, and $q = 0.9$, respectively, with $\theta = 90^\circ$. The inset panels include the $B = 0$ case for comparison, while the main panels provide a zoomed-in view to highlight the effect of the magnetic field. The plots demonstrate that increasing $B$ leads to a consistent decrease in $\Omega_p$, indicating that the magnetic field acts as a damping factor, suppressing spin precession in the weak-field limit.}

\end{center}
\end{figure}

 We now examine how the magnetic field influences spin precession in the weak-field regime, where $r \gg M$. This behavior is illustrated in Fig.~\ref{larger}, where we plot $\Omega_p$ as a function of $r$ for a BH. The main panel provides a zoomed-in view to highlight the effect of the magnetic field, while the inset panel includes the $B = 0$ case for comparison. As seen in panels (a), (b), and (c), for all values in the range $0 < q < 1$, increasing the magnetic field strength consistently leads to a reduction in the spin precession frequency, $\Omega_p$. This suggests that the external magnetic field acts as a damping factor, suppressing the precession frequency. The inset panel further confirms this trend: even when $B=0$, $\Omega_p$ decreases monotonically with increasing $r$. The presence of a magnetic field accelerates this decline, showing that the magnetic field significantly suppresses spin precession in the weak-field limit. This behavior is observed for all values of $\theta$ in the range $0 < \theta < \pi/2$, and for all values of $q$, except $q = 0.5$, which corresponds to ZAMO. While this behavior is explicitly illustrated for a BH in Fig.~\ref{larger}, we find that the same trend holds for NaS across the entire range $0 < q < 1$ and for all values of $\theta$, again with the sole exception of the ZAMO case.

\begin{figure}[h!]
\begin{center}
\subfigure[$a_*=0.9$, $q=0.5$, $\theta=10^\circ$]{
\includegraphics[width=2in,angle=0]{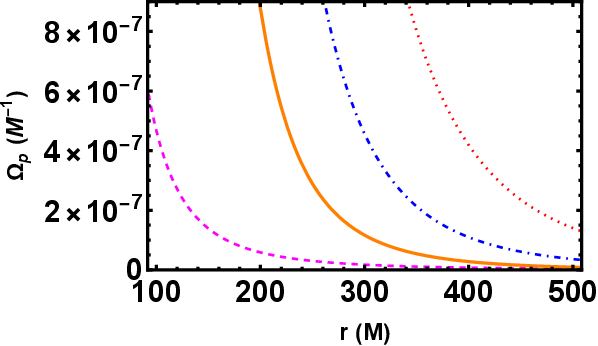}}
\subfigure[$a_*=0.9$, $q=0.5$, $\theta=20^\circ$]{
\includegraphics[width=2in,angle=0]{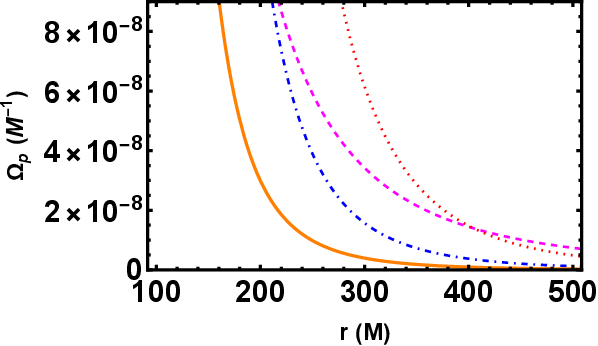}}
\subfigure[$a_*=0.9$, $q=0.5$, $\theta=30^\circ$]{
\includegraphics[width=2in,angle=0]{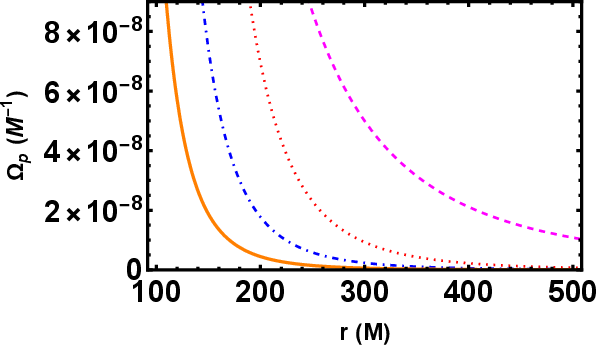}}
\includegraphics[width=3in,angle=0]{legend.eps}
\caption{\label{ZAMOlarger} $\Omega_p$ (in $M^{-1}$) is plotted as a function of $r$ (in $M$) in the weak-field regime ($r \gg M$) for a magnetized Kerr black hole. Panels (a), (b), and (c) correspond to $\theta = 10^\circ, 20^\circ,$ and $30^\circ$, respectively, with spin parameter $a_* = 0.9$ for ZAMO ($q = 0.5$). The plots illustrate that at lower inclination angles, the ZAMO frequency exhibits an anomalous trend where the magnetic field enhances rather than suppresses precession, in contrast to the general behavior discussed earlier. This effect diminishes as $\theta$ increases, aligning with the expected suppression of $\Omega_p$ due to the magnetic field.}
\end{center}
\end{figure}

For ZAMO $q=0.5$, the behavior is more nuanced. In the weak-field limit and at lower values of $\theta$ (closer to the polar axis), we observe an interplay between the GEM field and the external magnetic field, which disrupts the general damping trend. As we move towards the equatorial plane, this interplay fades, and the magnetic field begins to suppress $\Omega_p$ more effectively. This behavior is illustrated in Fig. \ref{ZAMOlarger}, where panels (a), (b), and (c) correspond to $\theta = 10^\circ, 20^\circ,$ and $30^\circ$, respectively. At smaller angles, the magnetic field has a relatively weaker influence, but as $\theta$ increases, its suppressive effect on spin precession becomes more evident. This peculiar behavior is unique to the ZAMO case and does not occur for other values of $q$, where the damping effect remains consistent across all $\theta$. Notably, while Fig. ~\ref{ZAMOlarger} explicitly shows this behavior for BH, the same trend persists for NaS in the ZAMO case at smaller values of $\theta$.

\vspace{2mm}

In summary, for all values of $\theta$ and $q$, we find that the magnetic field dampens the spin precession in the weak-field limit—except in the case of ZAMO ($q = 0.5$) at small $\theta$, where an interplay between the GEM field and the magnetic field temporarily alters this trend. This complete behavior is clearly illustrated in both Fig.~\ref{larger} and Fig.~\ref{ZAMOlarger}, for BH, and the same holds for NaS.

\section{\label{sec7} Lense-Thirring Precession in Magnetized Kerr Spacetime}

In this section, we discuss the LT precession of a test gyroscope due to the frame-dragging effects of the magnetized Kerr spacetime. To derive the LT precession frequency, we consider a test gyroscope attached to an observer in a stationary, axisymmetric spacetime. The general expression for the spin precession frequency vector, given in Eq.~\eqref{Omegavecp}, reduces to the LT precession frequency when the angular velocity $\Omega$ is set to zero. Applying this condition, we obtain the LT precession frequency \cite{Chakraborty2014,CCDKNS,StraumannGR}

\begin{align}
\vec{\Omega}_{\text{LT}} = \frac{1}{2\sqrt{-g}} \left[- \sqrt{g_{rr}} \left( g_{0\phi,\theta} - \frac{g_{0\phi}}{g_{00}} g_{00,\theta} \right) \hat{r}
+  \sqrt{g_{\theta\theta}} \left( g_{0\phi,r} - \frac{g_{0\phi}}{g_{00}}  g_{00,r} \right) \hat{\theta}\right].
\label{LTgeneral}
\end{align}

This result is valid only outside the ergoregion because at ergoregion $g_{00}$ goes to zero and $\vec{\Omega}_{\text{LT}}$ diverges at erogoregion. By substituting the metric components of the magnetized Kerr spacetime into Eq. \eqref{LTgeneral}, we obtain the LT precession frequency of a gyroscope

\begin{align}
    \vec{\Omega}_{\text{LT}} \approx \left( 2aMr\cos\theta \frac{\sqrt{\Delta}}{\Sigma^{3/2} (\Sigma-2Mr)} \hat{r} - aM\sin\theta \frac{(\Sigma-2r^2)}{\Sigma^{3/2}(\Sigma-2Mr)} \hat{\theta} \right) + B^2 \vec{\mathcal{F}} (r,\theta, M, a) + \mathcal{O}(B^3).
    \label{magkerrlt}
\end{align}

\begin{figure}[!h]
\begin{center}
\subfigure[~BH with $a_*=0.8$, $\theta=90^\circ$]{
\includegraphics[width=3in,angle=0]{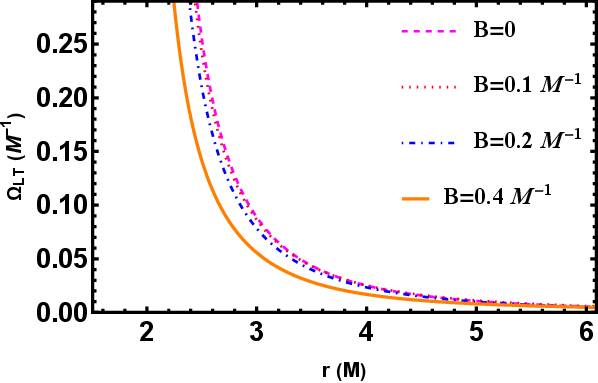}} 
\subfigure[~NaS with $a_*=2$, $\theta=45^\circ$]{
\includegraphics[width=3in,angle=0]{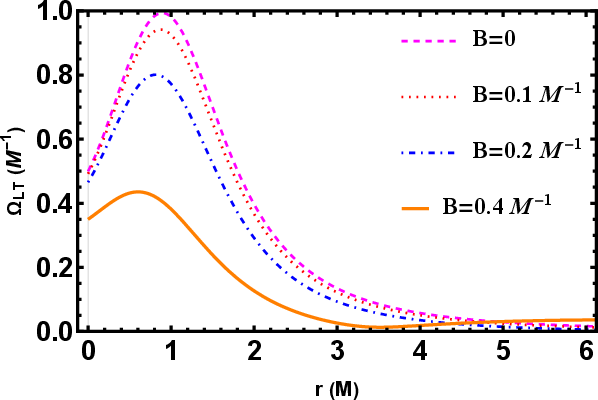}} 
\subfigure[~BH with $a_*=0.8$, $\theta=45^\circ$]{
\includegraphics[width=3in,angle=0]{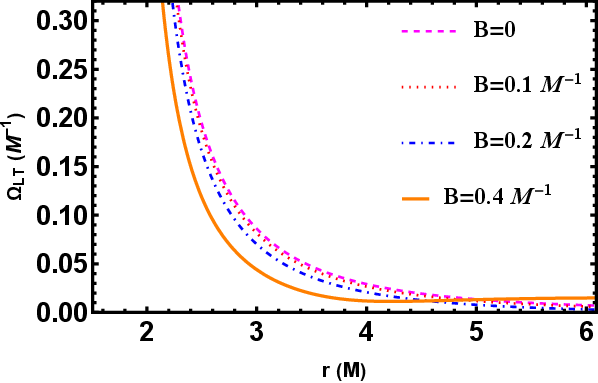}}
\subfigure[~NaS with $a_*=2$, $\theta=30^\circ$]{
\includegraphics[width=3in,angle=0]{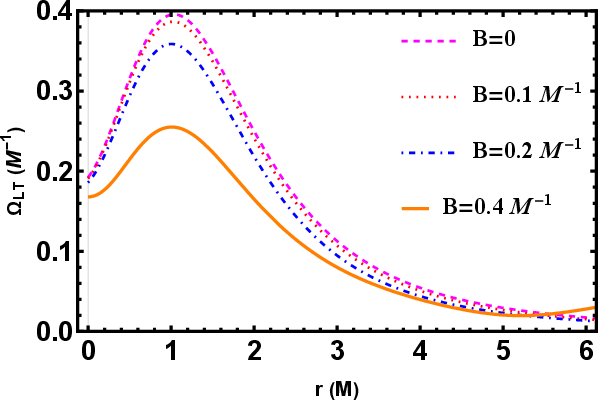}} 
\subfigure[~BH with $a_*=0.8$, $\theta=0$]{
\includegraphics[width=3in,angle=0]{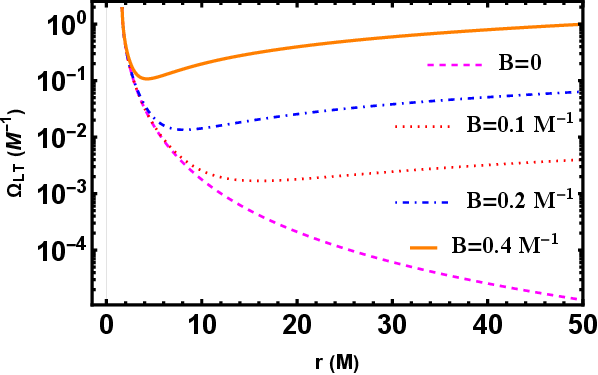}}
\subfigure[~NaS with $a_*=2$, $\theta=0$]{
\includegraphics[width=3in,angle=0]{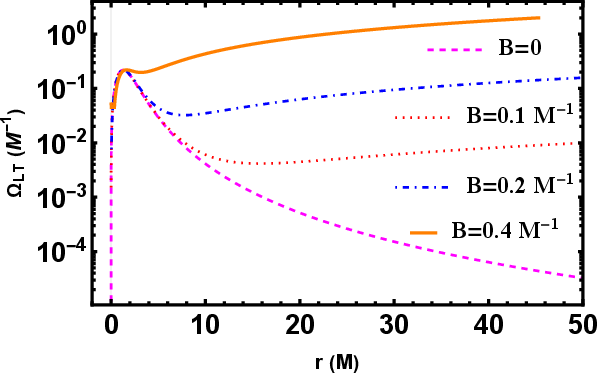}}
\caption{\label{LTMagKerr} $\Omega_{\text{LT}}$ (in $M^{-1}$) versus $r$ (in $M$) for a BH (left panels) with $a_*=0.8$ and NaS (right panels) with $a_*=2$ is shown for various values of $B$ and $\theta$. The plots clearly shows that $\Omega_{\text{LT}}$ increases with $B$ along the polar axis in contrast to as one moves toward the equatorial plane. There is an interplay between GM field and $B$ field in $0<\theta \lesssim \pi/2$. See Sec. \hyperref[sec7]{VII} for details.}
\end{center}
\end{figure}
It is evident that the first term in $\vec{\Omega}_{\text{LT}}$ corresponds to the Kerr spacetime and matches the result of  \cite{Chakraborty2014}. The second term, $\vec{\mathcal{F}}$ represents the leading-order contribution of the magnetic field to the LT precession \footnote{As before, given the length of the full expressions for $\vec{\Omega}_\text{LT}$ in the magnetized Kerr spacetime, we present only the leading-order contribution of the magnetic field to LT precession. Here. we present the graphical analysis to more intuitively illustrate their physical implications. The full analytical expressions are available upon request.}.

\vspace{2mm}

To investigate the effects of the magnetic field on LT precession in the strong-field regime, we plot the modulus of $\vec{\Omega}_{\text{LT}}$, given by Eq. \eqref{magkerrlt}, for a BH with $a_*=0.8$ and a NaS with $a_*=2$ for various values of $B$. From Fig. \ref{LTMagKerr}, panels (a), (c) and (e) illustrate that the LT precession frequency of a BH and panels (b),(d),(f) that of NaS. It is evident that the LT precession frequency of these test gyroscopes in the case of a BH, diverges at the ergoregion well before approaching $r=0$. However, in a NaS spacetime, gyroscopes can be positioned arbitrarily close to $r=0$ across a wide angular range, with their LT precession frequency remaining finite and well-defined throughout. It must be noted that LT precession for NaS diverges only at the ring singularity ($r=0,\theta=90^\circ$). Furthermore, as shown in panels (e) and (f) of Fig.~\ref{LTMagKerr}, along the polar axis ($\theta = 0$), the LT precession frequency increases with increasing value of $B$ for both the BH and NaS. It shows the opposite behavior along all other directions, as illustrated in panels (a-d) of Fig.~\ref{LTMagKerr}. This behavior arises due to the orientation of the magnetic field. Since the field is applied perpendicular to the equatorial plane, its interaction with frame-dragging effects is strongest along the polar axis. As the observer moves away from the pole toward the equatorial plane, the influence of the magnetic field on LT precession weakens, this interplay between the GM field and the magnetic field is illustrated in Fig.~\ref{LTMagKerr}.  

\subsection{Special Case: $r=0$}

As discussed in Sec. \ref{sec4}, our derived expression for $\vec{\Omega}_p$ is not valid at the ring singularity $(r=0, \theta=\pi/2)$. However, we can examine its limiting behavior near this point, particularly within the region $r \rightarrow 0$ for $0 \leq \theta <90^\circ$. Figure~\ref{LTMagKerr} demonstrates that while the LT precession frequency diverges at the ergoregion for BHs, it remains finite and well-behaved across a broad angular range in the case of a NaS spacetime. This distinction motivates a closer investigation of the behavior near $r=0$ as a special case. At this limit, Eq. \eqref{magkerrlt} reduces to 

\begin{align}
    \vec{\Omega}_{\text{LT}} \big|_{r=0} \approx \left( -\frac{1}{64} a N_1 B^4 M^2 \sec ^4\theta \right) \ \hat{r} + \left( -\frac{M}{a^2} \tan\theta \sec^2\theta - \frac{1}{32} N_2 M B^4 \right) \hat{\theta},
    \label{LTr0}
\end{align}

where

\begin{align}
    N_1 &=31 \cos 2 \theta +10 \cos 4 \theta +\cos 6 \theta +38, \nonumber \\
    N_2 &= \left(3 a^2+4 M^2\right) \sin 2 \theta -2 \left(\left(a^2+36 M^2\right) \sec ^2\theta -6 a^2+4 M^2 \sec ^4\theta -8 M^2\right) \tan \theta.
\end{align}

Thus we note that $\Omega_{\text{LT}}$ remains finite for $0\leq\theta<90^\circ$ at $r=0$. However, Eq. \eqref{LTr0} diverges at $\theta=90^\circ$ precisely on the location of ring singularity. If we set $B=0$, Eq. \eqref{LTr0} reduces to

\begin{align}
    |\Omega_{\text{Kerr LT}}| = \frac{M}{a^2} \tan\theta\sec^2\theta,
    \label{LTr0Kerr}
\end{align}

is consistent with previous calculations in Ref. \cite{CKJ},
as evident from Eqs. (6) and (7) therein (see also Sec. IV A of Ref. \cite{CCDKNS}).

\subsection{Special Case: Slow Rotation and Weak Field Approximation}

Another particularly important limiting case involves the LT precession frequency $\vec{\Omega}_{\text{LT}}$ under the assumptions of slow rotation and weak gravitational and magnetic fields. Specifically, for a spacetime with slow rotation ($a/M \ll 1$), weak gravity ($r \gg M$), and a weak magnetic field ($B \ll M^{-1}$), the general expression in Eq. \eqref{magkerrlt} simplifies to   

\begin{align}
\vec{\Omega}_{\text{LT}} (r,\theta) \approx \frac{aM}{r^3}[2 \cos\theta \hat{r} + \sin\theta \hat{\theta}] + \frac{aMB^2 \sin^2\theta}{r} \left[ -\frac{7}{2} \cos\theta \hat{r} + \frac{5}{4} \sin\theta \hat{\theta} \right].
\label{magkerrltweak}
\end{align}  

The first term in the above expression corresponds exactly to the standard Lense-Thirring precession frequency for the slowly-spinning Kerr BH in weak field approximation, which falls off as $ 1/r^3 $ \cite{Chakraborty2014,Hartle_2021,IorioLTSS}. The second term is the correction due to the presence of the magnetic field, which modifies the Lense-Thirring precession and introduces an additional contribution that falls off as $ 1/r $. 

\vspace{2mm}
A key distinction between these two terms lies in their radial dependence. In the standard Kerr case, the Lense-Thirring precession frequency decreases as $ 1/r^3 $, reflecting the rapid weakening of frame-dragging effects with distance. However, in the presence of an external magnetic field, the correction term decreases more slowly, as $ 1/r $. This suggests that while the influence of the magnetic field may be subdominant near the BH (where GM effects are strong), at sufficiently large distances, its contribution diminishes at a slower rate compared to the purely gravitational effect.  Physically, this behavior arises because the Kerr frame-dragging effect is primarily dictated by the gravitomagnetic field of the spinning mass, which decays rapidly due to its dipole nature. In contrast, the magnetic field-induced term originates from the interaction between the spin of the black hole and the external field, leading to a more gradual falloff. This implies that, in astrophysical scenarios where magnetic fields are present, their influence on precession may persist over larger distances than previously expected in the absence of a magnetic field.

\section{\label{geo} Non-vanishing spin precession in the non-rotating spacetime} 

It is interesting to see that the exact expression (equivalent to Eq. \ref{Spinmagkerr}) of spin precession  does not vanish for $a \rightarrow 0$. It reduces to 

\begin{align}
   \Big| \vec{\Omega}_p\Big|_{a\rightarrow 0,\theta \rightarrow \pi/2} 
    = \Omega\frac{64(-3B^2r^3+5B^2r^2M+4r-12M)}{(r-2M)(4+B^2r^2)^4-256r^3\Omega^2}
     \label{MagSch}
\end{align}
in the equatorial plane (i.e., $\theta=\pi/2$).
In Eq. (\ref{MagSch}), $\Omega$ can take any value ensuring $u$ remains timelike. This implies that a gyroscope moving in a non-rotating spacetime, despite being static, undergoes precession \cite{ccwormhole}. We stress that Eq. \eqref{MagSch} is an exact result, derived without any approximations. Now, if the gyroscope follows a circular geodesic, $\Omega$ must correspond to the Kepler frequency ($\Omega_{\text{Kep}}$). Moreover,  for $a \rightarrow 0$, Eq. \eqref{kerrmagm} reduces to the magnetized Schwarzschild spacetime \cite{CCGLP}.
The exact Kepler frequency for the magnetized Schwarzschild spacetime is given by Eq. (12) of \cite{CCGLP}:  
\begin{align}
    \Omega_{\text{Kep}} = \frac{(4+B^2r^2)^2}{16} \sqrt{\frac{M (4-3 B^2 r^2)+2 B^2 r^3}{r^3 (4-B^2 r^2)}}. \label{KepMagSch}
\end{align}
Substituting Eq. \eqref{KepMagSch} into Eq. \eqref{MagSch} yields  
\begin{align}
 \Omega_1 \equiv   \Omega_p \big|_{\left(a \rightarrow 0,~\theta \rightarrow \pi/2,~\Omega \rightarrow \Omega_{\text{Kep}}\right)} =  \frac{4}{\left(B^2 r^2+4\right)^2} \ \sqrt{\frac{(4-B^2 r^2) \ (4M-3 B^2 M r^2+2 B^2 r^3)}{r^3}}.
    \label{omegapkep1}
\end{align}

\begin{figure}[H]
\begin{center}
\includegraphics[width=3.4in,angle=0]{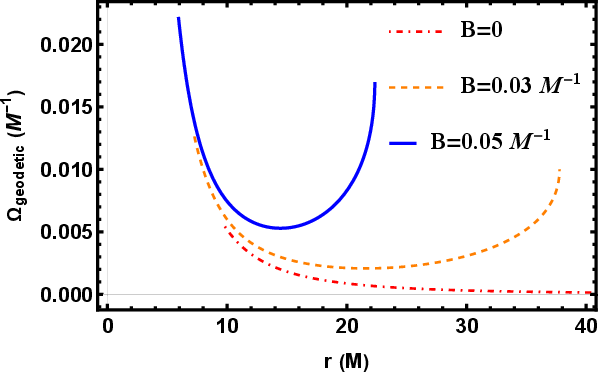}
\caption{\label{fig:Geodetic} $\Omega_{\text{geodetic}} (M^{-1})$ as a function of $r (M)$ for various values of $B (M^{-1})$. The curves are constructed from ISCO to OSCO. As $r$ increases, the magnetic field's influence becomes stronger, causing the geodetic precession to increase. See Sec. \ref{geo} for details.}
\end{center}
\end{figure}

This expression gives the precession frequency in the Copernican frame, measured with respect to proper time $\sigma$. The proper time $\sigma$, defined in the Copernican frame, relates to the coordinate time $t$ via  
\begin{align}
    d\sigma =\frac{\left(4+B^2 r^2\right)}{4} \sqrt{\frac{12M-4r-5M B^2 r^2+3B^2r^3}{r(B^2 r^2-4)}} dt.
\end{align}
The precession frequency in the coordinate basis $\Omega_2$ is then  

\begin{align}
    \Omega_2= \frac{1}{r^2(4+B^2r^2)} \sqrt{M \left(4-3 B^2 r^2\right)+2 B^2 r^3} \sqrt{M \left(5 B^2 r^2-12\right)-3 B^2 r^3+4 r}.
\end{align}
To determine the frequency associated with the change in the spin vector's angle over one full revolution around the central object, we compute the difference between $\Omega_2$ and $\Omega_1$ \cite{Hartle_2021}, and we get

\begin{align}
  \Omega_{\text{geodetic}} &= \sqrt{\frac{(4-B^2r^2)(4M-3B^2Mr^2+2B^2r^3)}{r^3}} \left(\frac{4}{(4+B^2r^2)^2}-\frac{1}{(4+B^2r^2)}\sqrt{\frac{12M-4r-5B^2Mr^2+3B^2r^3}{r(B^2r^2-4)}} \right) \nonumber \\
  \quad &\approx \sqrt{\frac{M}{r^3}} \left(1- \sqrt{1-\frac{3 M}{r}}\right)   \label{geodeticmagsch}
  \\
     &\quad + \frac{B^2\left [r^2 \left ( 1- \sqrt{1-\frac{3M}{r}} \right ) - 7Mr \left ( 1- \sqrt{1-\frac{3M}{r}} \right ) - 2M^2 \left (5 \sqrt{1-\frac{3M}{r}} - 6  \right )  \right ]}{4 (1-\frac{3M}{r}) \sqrt{Mr} } + \mathcal{O}(B^3).
      \label{geodeticmagscha}
\end{align}
Eq. \eqref{geodeticmagsch} is identified as the geodetic precession ($\Omega_{\text{geodetic}}$) frequency. We emphasize that Eq. \eqref{geodeticmagsch} is exact, with no approximations applied to $B$, whereas Eq. \eqref{geodeticmagscha} is approximated upto $\mathcal{O}(B^3)$. To verify consistency, we take the limit $B \to 0$ in Eq. \eqref{geodeticmagsch}, obtaining  

\begin{align}
     \Omega_{\text{geodetic}} \Big|_{B=0} = \sqrt{\frac{M}{r^3}} \left(1-\sqrt{1-\frac{3M}{r}} \right),
\end{align}
thus recovering the well-known result of \cite{SakinaSch} for Schwarzschild spacetime, confirming the validity of our derivation. In Fig. \ref{fig:Geodetic}, we plot $\Omega_{\text{geodetic}} (M^{-1})$ as a function of $r (M)$ for different values of $B (M^{-1})$. The curves are constructed between the Innermost Stable Circular Orbit (ISCO) and the Outermost Stable Circular Orbit (OSCO) \footnote{Notably, OSCOs do not exist in Schwarzschild or Kerr spacetime. The reason for the appearance of an OSCO in the magnetized Schwarzschild case is straightforward: as $r \to \infty$, the spacetime asymptotically approaches the Melvin universe, meaning the magnetic field remains non-vanishing even at large distances. Consequently, an additional stability condition arises, leading to the existence of OSCO.}. The ISCO radius ($r_I$) is determined by the expression \cite{CCGLP}:  

\begin{align}  
    &12 B^6 r_I^8 - 37 B^6 M r_I^7 + (30 B^6 M^2 - 48 B^4) r_I^6 + 204 B^4 M r_I^5 + (128 B^2 - 200 B^4 M^2) r_I^4 - 624 B^2 M r_I^3 \nonumber \\  
    &+ 672 B^2 M^2 r_I^2 + 64 M (r_I - 6M) = 0,  
    \label{isco}  
\end{align}  

while the OSCO radius ($r_O$) is given by the expression \cite{GaltsovBHMag}
\begin{align}
    r_{\text{O}} = \frac{2}{\sqrt{3} B}.
    \label{osco}
\end{align}

For $B=0$, the OSCO does not occur, and the geodetic precession increases near the ISCO, as expected due to the stronger curvature in this region. However, for non-zero magnetic fields ($B \neq 0$), we observe a more nuanced behavior: the geodetic precession initially increases near the BH, then gradually decreases, and finally rises again at larger $r$. This phenomenon can be understood as a tug-of-war between the gravitoelectric field and the magnetic field. Near the BH, the high curvature dominates, enhancing the geodetic precession. At intermediate distances, the effects of the GE field and the magnetic field balance each other, resulting in a local minimum. As $r$ increases further, the magnetic field's influence becomes stronger, causing the precession to increase thus demonstrating that the magnetic field supports geodetic precession. This behavior is clearly illustrated in Fig.~\ref{fig:Geodetic}. 

\vspace{2mm}

In the limiting case of weak gravity ($r \gg M$) and a weak magnetic field, the geodetic precession frequency $\Omega_{\text{geodetic}}$ reduces to

\begin{align}
\Omega_{\text{geodetic}}\Big|_{r \gg M} &\approx \frac{3 M^{3/2}}{2 r^{5/2}} + \frac{3}{8} \sqrt{Mr} \left(1 - \frac{23}{12} \frac{M}{r}\right) B^2.
\label{geol}
\end{align}
This result demonstrates that, even in the weak-field regime, the magnetic field has a non-negligible influence on spin precession at large distances. While the pure gravitoelectric contribution ($\propto r^{-5/2}$) decays rapidly, the magnetic contribution remains significant far from the source.

\subsection{Special case:  {\it Geodetic precession} in the Melvin’s magnetic Universe}
One intriguing feature that emerges is, the non-vanishing geodetic precession for $M \rightarrow 0$. This implies that Eq. (\ref{geodeticmagsch}) reduces to 

\begin{align}
  \Omega_{\text{geodetic}}|_ {M \rightarrow 0} &=  \frac{4\sqrt{2}B}{\left(4+B^2 r^2\right)^2} \sqrt{4- B^2 r^2} \left[1-\left(1+\frac{B^2 r^2}{4}\right) \sqrt{\frac{4-3 B^2 r^2}{4-B^2 r^2}}\right]
\label{mg} 
  \\
  &\approx \frac{5 B^5 r^4}{32 \sqrt{2}} + \mathcal{O}(B^6)
\end{align}
in the limit of $M \rightarrow 0$.
Note that $M\rightarrow 0$ with a non-zero magnetic field implies the Melvin magnetic universe \cite{melvinverse} which is an exact solution of the Einstein-Maxwell equations.
%, where the magnetic pressure balances the gravitational attraction. 
The non-vanishing geodetic precession in the Melvin's spacetime arises solely due to the magnetic curvature arisen from the (magnetic-)energy density ($\sim B^2$) of the real magnetic field that contributes to the gravitational mass of the spacetime \cite{Bonnor_1960,chakraborty2023geometric} in this case.

\section{\label{sec8} Summary and Discussion}

This paper has presented several important and insightful results, which we summarize as follows:

\begin{enumerate}
    \item Considering the exact solution of the magnetized Kerr spacetime, we have derived the exact expression of spin precession of a test gyroscope which are valid for the magnetized Kerr BH as well as magnetized Kerr NaS.
    
    \item Our result is applicable from the weak magnetic field to the ultra strong magnetic field (i.e., 0 to any arbitrary value of $B$) which can even distort the original Kerr geometry.
    
    \item The precession frequencies of the spin of test gyros attached to timelike stationary observers exhibit distinct behaviors for BHs and NaSs. For BHs, the frequencies are finite both inside and outside the ergoregion but become arbitrarily large as the gyro approaches the horizon ($r \sim r_+$) in any direction ($0 < \theta \leq \pi/2$). In contrast, for a NaS, the frequencies remain finite and regular even at $r \rightarrow 0$ for all $\theta \nsim \pi/2$. These results suggest that gyroscope precession frequencies can serve as an effective tool to distinguish BHs from NaSs, offering a potential observational signature to identify the nature of the central compact object. 
  
    \item For gyroscopes placed increasingly closer to a BH or a NaS in the equatorial plane, their spin precession frequencies become arbitrarily large in both cases, but in distinct ways: the frequencies increase sharply near the event horizon for a BH, whereas they grow increasingly large as $r \rightarrow 0$ for a NaS.
    
    \item We have shown that the spin precession frequency stays finite in the limit of the approach to the horizon for a ZAMO ($q=0.5$), while it becomes arbitrarily large in all other cases.
 
    \item Our analysis reveals a distinct tug-of-war behavior in the precession frequency of gyro in magnetized Kerr spacetime, arising from the competing influences of the GEM field and the external magnetic field. This interplay is particularly pronounced at intermediate distances from the central object, where neither field fully dominates.
    
    \item In the weak-field regime ($r \gg M$), we find that the presence of an external magnetic field generally acts as a damping agent, reducing the spin precession frequency $\Omega_p$ across all values of inclination angle $\theta$ and observer parameter $q$, except for the ZAMO case ($q = 0.5$) at low $\theta$, where an interplay between the magnetic and GM fields temporarily disrupts this trend.
    
    \item  We observe that the LT precession frequency remains finite in regions outside the ergoregion but becomes very high as an observer approaches the ergoregion boundary in the case of a BH. In contrast, for the NaS, the LT precession frequency stays finite across a broad angular domain except at the ring singularity. Notably, along the polar axis ($\theta \rightarrow 0$), the precession frequency shows a distinct behavior—it increases with the magnetic field strength $B$, unlike in other directions where this trend is not observed.
    
    \item We derive the LT precession for a slowly spinning BH in the weak gravity limit and find that magnetic fields introduce a correction to the standard Kerr term. While the Kerr contribution falls off as $r^{-3}$, the magnetic correction decreases more slowly ($\propto r^{-1}$), suggesting that magnetic effects can dominate at large distances.
    
    \item We obtain the exact geodetic precession frequency for a test gyroscope in magnetized Schwarzschild spacetime. For $B\neq 0$, geodetic precession increases near the BH due to strong curvature, then decreases at intermediate distances as GE and magnetic effects balance, and increases again at larger $r$ where the magnetic field dominates, thus suggesting the magnetic field increases the geodetic precession.
    
    \item In the weak-field regime ($r \gg M$) and under a weak magnetic field, the geodetic precession frequency acquires a magnetic correction that scales as $\propto r^{1/2}$, indicating that magnetic fields can noticeably affect spin precession even at large distances from the central object. Indeed, Eq. (\ref{geol}) shows that $\Omega_{\rm geodetic} \propto B^2\sqrt{Mr}$, implying that even a small magnetic field, if extended over a sufficiently large region, can lead to a substantial precession effect. This could have important implications for galactic astrophysics, in future.

    \item Remarkably, in the $M \to 0$ limit corresponding to the Melvin magnetic universe, a finite geodetic precession arises purely due to magnetic curvature. This emphasizes that the magnetic fields can induce spin precession even in the absence of mass, highlighting the independent role of magnetic geometry in influencing spin precession.
\end{enumerate}

%\vspace{2mm}

The magnetized Kerr spacetime employed in our analysis is not asymptotically flat. The magnetic field is assumed to be asymptotically uniform following the approach of \cite{CCPenrose}. Despite this limitation, the metric remains an exact electrovac solution of the Einstein–Maxwell equations. It is also standard practice in the literature to model the magnetic field around collapsed objects (BH or NaS) as uniform, since such configurations are more analytically tractable and have been successfully used to study a wide range of astrophysical phenomena (e.g., \cite{Dadhich, CCFaraday, CCPenrose, Chatterjee_2017, Magkerrmeissner} and so on). Moreover, the absence of reliable measurements or precise knowledge of magnetic field geometries near compact objects \cite{punslyBHGH, Davidengine} justifies this simplifying assumption. Nonetheless, future work could aim to refine the spin precession formalism by incorporating spatially varying magnetic fields, which would provide a more realistic description of such astrophysical environments.
{While our analysis reveals rich and distinct spin precession behaviors in the magnetized Kerr spacetime, we acknowledge the theoretical and interpretive challenges that come with employing such models. In particular, we consider a Kerr BH/NaS immersed in an external uniform magnetic field — a widely used but idealized setup that facilitates exact analytical treatment. However, this uniform field configuration does not arise from realistic astrophysical collapse and leads to a non-asymptotically flat spacetime that asymptotically approaches the Melvin universe. From a more mathematical point of view, differently, the spacetime can have a certain interest. These features, while mathematically consistent within the Einstein-Maxwell framework, limit the direct physical interpret ability of the results in realistic settings. Nevertheless, this construction serves as a valuable testbed to explore how external magnetic fields influence gyroscopic spin precession, especially through their coupling with the GEM structure of the spacetime. Our goal here is not to assert a fully realistic model of strong gravity near compact objects, but rather to identify qualitative and potentially observable effects that may persist in more complete descriptions. A more detailed analysis incorporating self-consistent sources of the magnetic field or numerical GR simulations may be necessary to bridge the gap between theoretical insights and astrophysical observables. Despite these limitations, we have outlined several potential astrophysical scenarios \cite{CCPenrose, CCFaraday, BHshield, CCGLP, adarsha2025,adarsha2}, where this formulation could be relevant, as discussed below.

Although the magnetic field strengths in the X-ray corona of the black hole Cygnus X-1 have been estimated to reach up to $10^7$ G \cite{santomnrs}, the current emission models for Sagittarius A* and M87* suggest significantly lower field strengths, with typical values of $B \sim 30$–$100$ G \cite{eatough2013strong} and $B \sim 1$–$30$ G \cite{EHT7,EHT8}, respectively. To date, ultra-strong magnetic fields of the order of Eq.~\eqref{bmax} have not been observationally confirmed in the vicinity of any astrophysical black hole. The existence of such fields remains an open question in contemporary astrophysics. Even if such extreme magnetic fields do exist near collapsed objects, they may evade detection due to possible magnetic shielding effects \cite{BHshield}. 
Nonetheless, the predictions presented in this work could be tested if future missions such as ngEHT \cite{ngEHT}, BHEX \cite{BHEX} were to uncover evidence of ultra-strong magnetic fields in the vicinity of compact objects. For instance, a potential scenario involves a magnetar companion near Sagittarius A*, such as SGR J1745–29, which may provide magnetic fields as high as $B = 1.6 \times 10^{14}$ G \cite{kennea2013swift,mcgill}. Another potential scenario involves ultra-strong magnetic fields of the order of $10^{20}$ G, which may have originated in the early Universe (see \cite{GRASSO1,GRASSO2} and references therein), potentially immersing primordial black holes (PBHs) in such fields. It is also conceivable that primordial naked singularities (PNaSs) \cite{JoshiPNaS} formed in the early Universe or low-mass NaSs formed via dark core collapse \cite{CBJDarkcore} could be embedded in similarly strong magnetic environments. In such contexts, our results on spin precession in the presence of ultra-strong magnetic fields could find direct applicability.

\vspace{2mm}

In case of the above-mentioned magnetic shielding effect \cite{BHshield}, a test particle may be unable to follow its regular geodesic motion. Consequently, the formation of an accretion disk could be suppressed, and the orbital motion of surrounding stellar objects may be absent. It has been suggested \cite{BHshield} that although such isolated BHs do not emit detectable radiation, their gravitational field can still bend and focus light from background sources. Therefore, there remains a possibility of detecting them via gravitational lensing. We should emphasize it here that the magnetically shielded BHs could also be detected by studying the behavior of a test gyroscope or spinning particle. Specifically, we have shown in this paper that such a gyroscope exhibits a non-zero precession frequency in the vicinity of a BH immersed in an ultra-strong magnetic field. Since a test gyroscope does not follow geodesic motion \cite{CCDKNS}, it could potentially serve as a probe to detect such magnetically shielded collapsed objects.

\vspace{2mm}

Our exact result of spin precession could also be relevant for BHs formed via collapse induced by an endoparasitic BH (EBH) through the accumulation of dark matter within a neutron star \cite{CBJDarkcore}, or a white dwarf \cite{CCSubsolar}, magnetar, or similar compact object with a magnetic field \cite{CCPenrose, BHshield}. Additionally, it applies to BHs resulting from the merger of a BH with one or more magnetized neutron stars,  magnetars \cite{lyu83, east, BHshield}, or magnetized progenitors. In such scenarios, the magnetic field may not slide off the newly-formed BH, as demonstrated in \cite{lyu, lyu83} (but see \cite{brp}). The spin precession around these magnetized BHs can then be investigated using our exact formulation.
It has recently been proposed that (magnetized) white dwarfs might host EBHs at their centers that do not immediately destroy the host star. In certain models, the EBHs could be accessible  by the polar tunnels \cite{adarsha2025} developed in the said white dwarfs. One could hypothetically investigate the spin precession effects in such a scenario by lowering a test gyroscope along the polar tunnel into the near-horizon region of the EBH immersed in the strong magnetic field of the magnetized white dwarfs \cite{Bhattacharya_2022}. Similarly, the effect of magnetic fields on spin precession could also be measured hypothetically for an EBH formed inside a magneter \cite{adarsha2}. This could also enable the study of the gravitoelectromagnetic fields structure near a(n) (E)BH in a strong magnetized environment. 

\vspace{2mm}

One of the key features of our analysis is the exact expression for the Lense-Thirring precession frequency in the presence of an external magnetic field, derived by setting $\Omega = 0$ in the general spin precession formula. Our formulation can also be extended to compute the corresponding Lense-Thirring torque. It would therefore be interesting to investigate how this magnetically modified torque influences astrophysical phenomena such as the Bardeen–Petterson effect \cite{BPOG}. While previous studies have shown that the inner disk may not necessarily align with the BHs spin axis under certain viscous conditions \cite{banerjee2019}, our formulation opens up the possibility of examining whether the presence of magnetic fields can further affect the tilt evolution and warp alignment behavior \cite{CBMNRAS}. Geodetic precession effects on pulsar timing have been extensively studied, providing a precise observational test for gravitational theories (e.g., \cite{Hotan_2005, Konacki_2003}). Our exact analysis of geodetic precession in a magnetized Schwarzschild spacetime may offer a theoretical foundation for understanding how such precession influences pulsar timing in magnetized environments near compact objects. Geodetic precession effects on pulse timing have been extensively studied and provide a precise observational testbed for gravitational theories(e.g., \cite{Hotan_2005,Konacki_2003} and so on).
It was shown \cite{GMPulsar} that a rotating BH induces precession of a nearby pulsar’s spin axis due to geodetic and frame-dragging effects. Our expressions for geodetic and LT precession may find direct application in modeling pulsar spin dynamics near magnetized SMBHs. Specifically, these results could refine predictions of pulse arrival times and beam orientation changes in environments where strong-field gravity and electromagnetic interactions coexist.

{In this paper, we have studied the spin precession for both the Kerr BH and Kerr NaS. Although NaSs remain speculative due to their violation of the Cosmic Censorship Conjecture, we have included them in our analysis for both theoretical and observational reasons. Recent developments suggest it is premature to dismiss such configurations entirely. The Event Horizon Telescope Collaboration has discussed \cite{EHT2022SgrA} the possibility that Sagittarius A* could be described by a Joshi-Malafarina-Narayan \cite{Joshi2011Collapse} (JMN-1) type naked singularity. Similarly, both M87*, a supermassive collapsed object, and GRO J1655–40, a stellar-mass compact object in an X-ray binary system, have been proposed as potential NaS candidates \cite{GhasemiNodehi2021M87,CCGMmonople}. These discussions, while still debated, highlight the need for flexible theoretical frameworks that can accommodate a broader class of compact objects. In particular, a recent study \cite{DeliyskiNaS} demonstrates that reflective NaSs can produce distinctive observational signatures—such as enhanced central brightness and bright ring structures—potentially detectable by next-generation EHT arrays \cite{ngEHT}. Accordingly, we include NaS parameter regimes in our study not to advocate their physical realization, but to provide a more comprehensive understanding of how spin precession signatures may vary across different extreme geometries. This may prove useful for future efforts to observationally distinguish between BHs and more exotic alternatives in the presence of magnetic fields.

\vspace{2mm}

{\bf Acknowledgements:} The authors acknowledge the support of Manipal Academy of Higher Education.

\bibliography{ref}
\bibliographystyle{apsrev4-2}

\end{document}